\PassOptionsToPackage{sort&compress}{natbib}
\documentclass[times,review]{elsarticle}

\usepackage{multirow}

\usepackage{amssymb}
\usepackage{latexsym}

\usepackage{lineno}
\usepackage{amsmath}
\usepackage{bm}
\usepackage{subcaption} 
\usepackage{hyperref}
\usepackage{algcompatible} 
\usepackage{algorithm}
\usepackage{algpseudocode}

\usepackage[colorinlistoftodos]{todonotes}

\usepackage{siunitx}

\usepackage[titletoc,toc,title]{appendix}

\usepackage{placeins}
\usepackage{booktabs} 
\usepackage{graphicx}

\usepackage{graphicx}

\usepackage{tikz}
\usetikzlibrary{shapes}

\makeatletter
\def\BState{\State\hskip-\ALG@thistlm}
\makeatother

\makeatletter
\define@key{Gin}{splash_cylinder_crop}[]{\setkeys{Gin}{trim=25 35 260 0, clip,width=0.23\linewidth}}
\makeatother

\usepackage{graphicx}
\usepackage{environ}
\NewEnviron{myequation}{%
\begin{equation}
\scalebox{1.05}{$\BODY$}
\end{equation}
}

\begin{document}

	
	\begin{frontmatter}
		
	\title{Investigate the efficiency of incompressible flow simulations on CPUs and GPUs with BSAMR}%

    \author[1]{Dewen Liu}
    \address[1]{School of Civil Engineering, Southwest Jiaotong University, Chengdu, 610031, China}
    \author[2]{Shuai He}
    \address[2]{School of Power and Energy, Northwestern Polytechnical University, Xian, 710060, China}
    \author[3]{Haoran Cheng}
    \address[3]{Department of Computer Science, University of Michigan, N500 S State St, Ann Arbor, MI 48109}    
    \author[4]{Yadong Zeng\corref{cor1}}
    \ead{zengx372@utexas.edu}
    \cortext[cor1]{Corresponding author}
    \address[4]{Department of Computer Science, University of Texas at Austin, Austin, TX 78712, USA}

\begin{abstract}

Adaptive mesh refinement (AMR) is a classical technique about local refinement in space where needed, thus effectively reducing computational costs for HPC-based physics simulations. Although AMR has been used for many years, little reproducible research discusses the impact of software-based parameters on block-structured AMR (BSAMR) efficiency and how to choose them. This article primarily does parametric studies to investigate the computational efficiency of incompressible flows on a block-structured adaptive mesh. The parameters include refining block size, refining frequency, maximum level, and cycling method. A new projection skipping (PS) method is proposed, which brings insights about when and where the projections on coarser levels are safe to be omitted. We conduct extensive tests on different CPUs/GPUs for various 2D/3D incompressible flow cases, including bubble, RT instability, Taylor Green vortex, etc. Several valuable empirical conclusions are obtained to help guide simulations with BSAMR. Codes and all profiling data are available on GitHub.

\end{abstract}


\end{frontmatter}


\section{Introduction} \label{S:1}

Adaptive Mesh Refinement (AMR) is used in numerical simulations to enhance the resolution of specific regions while maintaining a low overall computational cost. Initially introduced in the 1980s~\cite{berger1984adaptive,berger1989local}, AMR has been widely adopted and refined by numerous researchers and practitioners across various disciplines. Examples include biomedical engineers who address heart valves~\cite{griffith2007adaptive,griffith2012immersed}, control and mechatronic engineers who address swimming fish~\cite{bhalla2013unified, bhalla2013forced, zeng2022subcycling}, electrical engineers who address electric magnetic field~\cite{fujimoto2006electromagnetic, balsara2001divergence, yao2021massively}, and mechanical and energy engineers who address ocean currents~\cite{santilli2015stratified}, atmospheric boundary layer~\cite{van2018towards}, wind turbines~\cite{sharma2024exawind, mullowney2021preparing}, and wave energy converters (WECs)~\cite{khedkar2021inertial, yu2013reynolds, yadong2020osm}.

One evident fact is that the use of adaptive meshes can significantly accelerate simulation speed and save more computational resources compared to uniformly refined grids throughout the entire domain~\cite{berger1984adaptive,almgren1998conservative}. Yet, there has been limited research that quantitatively analyzes how different parameters affect the computational efficiency of adaptive meshes. This work aims to fill this research gap by providing a comprehensive analysis of the impact of various parameters on the computational efficiency of AMR.

Adaptive meshes are commonly classified into two types, i.e.,~tree-based adaptive mesh~\cite{popinet2003gerris,popinet2009accurate} and block-structured adaptive mesh~\cite{berger1984adaptive,berger1989local,williamschen2013parallel}. In the tree-based adaptive mesh, a hierarchical tree-like structure is formed, where each cell is further divided into smaller cells, creating parent-child relationships between cells. On the other hand, the block-structured adaptive mesh does not individually partition specific cells; instead, it groups multiple cells as a patch for partitioning. In this study, we focus on the block-structured adaptive mesh refinement (BSAMR) method and utilize the open-source software Incompressible Adaptive Mesh Refinement (IAMR)~\cite{almgren1998conservative} to conduct a series of simulations of incompressible fluid flow. Compared to other open-source software for block-structured adaptive meshes, IAMR stands out for providing a versatile platform that allows simulations on both CPUs and GPUs. Various cases in the IAMR have been validated and integrated into the Continuous Integration (CI) and Continuous Development (CD) tests. Additionally, IAMR offers detailed profiling capabilities, which is a distinguishing feature as other adaptive mesh software may have limited profiling support for GPUs.

Numerous studies have integrated block-structured adaptive meshes into simulations of incompressible flows, primarily focusing on physics. They showed the adaptive grids' capability to accurately represent small-scale structures and elucidate their underlying mechanisms~\cite{balaras2009adaptive, bhalla2013unified, yu2013reynolds, griffith2007adaptive}. Yet, these investigations neglected to assess how adaptive grid algorithms and their associated parameters affect computational efficiency. This oversight can be attributed to the academic environment's emphasis on code correctness and physical accuracy over code optimization for speed, where understanding fluid dynamics mechanisms is prioritized over computational efficiency. In this work, we seek to fill the void by examining the impact of various parameters on the efficiency of adaptive grid-based simulations.

The structure of this article is as follows. Session~\ref{S:2} introduces the mathematical formulas of the incompressible fluid solver, the cycling method on the multilevel grid, and the open-source incompressible flow code and profiling data. Session~\ref{S:3} discusses the crucial parameters related to BSAMR studied in this article and qualitatively examines how these parameters affect the computational efficiency of block-structured adaptive grids. Session~\ref{S:4} describes the testing setup and Session~\ref{S:5} presents various test cases and conducts extensive testing on CPUs and GPUs, followed by a quantitative analysis of the computational results. Finally, Session~\ref{S:6} concludes the article and provides an outlook for future research directions.

\section{Mathematical formulation} \label{S:2}
\subsection{Projection-based Fluid Solver} \label{S:21}
We start with the incompressible Navier--Stokes equations for variable density flows,
\begin{linenomath*} \begin{equation}\label{eq:ns3}
\begin{aligned}
 \rho(\mathbf{x}, t) \left( \frac{\partial \mathbf{u}(\mathbf{x}, t)}{\partial t}+\nabla \cdot \mathbf{u}(\mathbf{x}, t) \mathbf{u}(\mathbf{x}, t) \right )&=-\nabla p(\mathbf{x}, t)+\nabla \cdot\left[\mu\left(\nabla \mathbf{u}(\mathbf{x}, t)+\nabla \mathbf{u}(\mathbf{x}, t)^{T}\right)\right]+\rho(\mathbf{x}, t) \mathbf{g}, 
\end{aligned}
\end{equation}\end{linenomath*}
\begin{linenomath*} \begin{equation}\label{eq:ns2}
\begin{aligned}
\nabla \cdot \mathbf{u}(\mathbf{x}, t) &=0,
\end{aligned}
\end{equation}\end{linenomath*}
where $p(\mathbf{x}, t)$, $\mathbf{u}(\mathbf{x}, t)$, and $\rho(\mathbf{x}, t)$, are the spatially and temporally varying pressure, fluid velocity, and density, respectively. Also, $\mu$ is the dynamic viscosity and $\mathbf{g}$ is the vector form of the gravitational acceleration.

To solve the above partial differential equations (PDEs), the canonical projection~\cite{chorin1967numerical} is applied to the semi-staggered mesh, in which fluid velocity, density, and scalar variables are located at the cell center and the pressure is located at the node center. The temporal and spatial discretizations of equations for single-level advancement are considered here. At the time $t^{n}$, the velocity $\mathbf{u}^{n}$, the density $\rho^{n}$, and pressure $p^{n-1/2}$ are known. The time step during the interval $[t^{n}, t^{n+1}]$ proceeds as follows.

\textbf{Step 1}: The density $\rho$, which is used to describe the two-phase interface, is updated by
\begin{linenomath*} \begin{equation}\label{eq:phi_advance}
\begin{aligned}
\frac{\rho^{n+1}-\rho^{n}}{\Delta t}+Q\left(\mathbf{u}^{n+\frac{1}{2}}_{\rm adv}, \rho^{n+\frac{1}{2}}\right)=0
\end{aligned}
\end{equation}\end{linenomath*}
where $Q\left(\mathbf{u}^{n+\frac{1}{2}}_{\rm adv}, \rho^{n+\frac{1}{2}}\right)$ is computed using the second-order Godunov scheme~\cite{almgren1998conservative,sussman1999adaptive, sverdrup2018highly, zeng2023consistent}.  The midpoint values of $\rho$ is then calculated as $\rho^{n+\frac{1}{2}} = (\rho^{n+1}+\rho^{n})/2$. Note this step can be omitted if $\rho$ is constant in the whole computational domain.

\textbf{Step 2}: The intermediate velocity ${\mathbf{u}}^{*,n+1}$ is solved semi-implicitly as
\begin{linenomath*}
\begin{linenomath*} \begin{equation} \label{eq:31}
\begin{aligned}
\rho^{n+\frac{1}{2}}\left(\frac{{\mathbf{u}}^{*,n+1}-\mathbf{u}^{n}}{\Delta t}+\bm{\nabla} \cdot \left(\mathbf{u}\mathbf{u} \right)^{n+\frac{1}{2}}\right)=-\bm{\nabla} p^{n-\frac{1}{2}}+
\frac{1}{2}\left(\bm{\nabla} \cdot \mu \bm{\nabla}{\mathbf{u}}^{*,n+1}+
\bm{\nabla} \cdot \mu \bm{\nabla} {\mathbf{u}}^{n}\right)+\rho^{n+\frac{1}{2}}\mathbf{g},
\end{aligned}
\end{equation}\end{linenomath*}
\end{linenomath*}
where the convective term $\bm{\nabla} \cdot \left(\mathbf{u}\mathbf{u} \right)^{n+\frac{1}{2}}$ is calculated using the same Godunov scheme as the density advection in \textbf{Step 1}.

\textbf{Step 3}: With the calculated intermediate velocity in \textbf{Step 2}, a level projection operator is applied to obtain the updated pressure $p^{n+1/2}$ and velocity ${\mathbf{u}}^{n+1}$fields. An auxiliary variable $\boldsymbol{V}$ is first calculated by
\begin{linenomath*}
\begin{linenomath*} \begin{equation}\label{eq:ns_lp1}
\boldsymbol{V} =  \frac{{\mathbf{u}}^{*,n+1}}{\Delta t} + \frac{1}{\rho^{n+1/2}} \bm{\nabla} p^{n-\frac{1}{2}}.
\end{equation}\end{linenomath*}
\end{linenomath*}
Then, $\boldsymbol{V}$ is projected onto the divergence-free velocity field to obtain the updated pressure $p^{n+1/2}$ via

\begin{linenomath*}
\begin{linenomath*} \begin{equation}\label{eq:ns_lp2}
L^{cc,l}_{\rho^{n+1/2}} p^{n+1/2} =  \bm{\nabla} \cdot \boldsymbol{V},  
\end{equation}\end{linenomath*}
\end{linenomath*}
where $L^{cc,l}_{\rho^{n+1/2}}p^{n+1/2}$ is the density-weighted Laplacian operator to $\bm{\nabla} \cdot (1/\rho^{n+1/2}\bm{\nabla} p^{n+1/2})$ on level $l$~\cite{almgren1998conservative,zeng2022aparallel}. Finally, the divergence-free velocity ${\mathbf{u}}^{n+1}$ on level $l$ is obtained as

\begin{linenomath*}
\begin{linenomath*} \begin{equation}\label{eq:ns_lp3}
{\mathbf{u}}^{n+1} = \Delta t \left(\boldsymbol{V} - \frac{1}{\rho^{n+1/2}} \bm{\nabla} p^{n+1/2}\right).
\end{equation}\end{linenomath*}
\end{linenomath*}
The projection is stable and appears to be well-behaved in various numerical tests~\cite{almgren1996numerical,rider1995approximate} and practical applications~\cite{sussman1999adaptive,martin2000cell}.

\subsection{Cycling method on the multiple levels} \label{S:22}
Two cycling methods are employed in this work to update variables on the multilevel grid. The first method, known as subcycling, involves advancing solutions on different levels with varying time-step sizes. It is assumed that the Courant-Friedrichs-Lewy (CFL) number remains constant across different levels, and the refinement ratio (denoted as $r$) between two consecutive levels~\cite{zeng2022parallel,zeng2019unified} is a fixed value. Consequently, the time-step sizes on levels $l$ and $l+1$ satisfy the relationship $\Delta{t^{l}}=r\Delta{t^{l+1}}$ for all $0 \leq l < l_{max}$. For a multilevel grid with $l_{max}=n-1$, we have $\Delta{t^{0}}=r\Delta{t^{1}}=r^{2}\Delta{t^{2}}=...=r^{n-1}\Delta{t^{n-1}}$ when utilizing the subcycling method. Conversely, the non-subcycling method, which involves all levels adopting the same time-step size as the finest level to prevent instability, resulting in $\Delta{t^{0}}=\Delta{t^{1}}=\Delta{t^{2}}=...=\Delta{t^{n-1}}$. Comparing the two methods, the subcycling method is more efficient than the non-subcycling method for advancing the solution from $t^{n}$ to $t^{n+1}$ due to its fewer substeps. However, the subcycling method necessitates time interpolation to address the temporal discrepancy among different levels~\cite{almgren1998conservative,nonaka2011three,bell1991efficient}, whereas the non-subcycling method eliminates the need for such time interpolation as all levels are synchronized at the same time instance.

In Section~\ref{S:5}, each simulation case uses $\Delta t_{0}$ as the time step on level 0. The grid spacings on level 0, denoted by $\Delta x_{0}$, $\Delta y_{0}$, and $\Delta z_{0}$, represent the spacing between grid points in the $x$-, $y$-, and $z$-directions, respectively. In the case of a multilevel grid, the grid spacings on level $l$ are determined by the relations $\Delta x_{l}=\Delta x_{0}/2^{l}$, $\Delta y_{l}=\Delta y_{0}/2^{l}$, and $\Delta z_{l}=\Delta z_{0}/2^{l}$, where $0 \leq l \leq l_{max}$. The determination of the time step size at the finest level, $\Delta t^{l_{max}}$, is based on considerations such as the CFL condition, gravity, and viscosity~\cite{zeng2022parallel,sussman1999adaptive,zeng2022subcycling,zeng2022numerical}.

\subsection{Open-source incompressible flow code and profiling data} \label{S:23}

While AMR has been extensively utilized in numerous simulations, it is necessary to investigate the influence of parameters on the computational efficiency of AMR and the optimization of its parameters. IAMR, a parallel and adaptive mesh refinement (AMR) code, incorporates subcycling in time and effectively solves the variable-density incompressible Navier-Stokes equations within complex geometries. IAMR is constructed upon the AMReX software framework~\cite{zhang2019amrex,zhang2020amrex}, which is a publicly accessible platform designed specifically for developing massively parallel block-structured adaptive mesh refinement (BSAMR) applications. The code supports hybrid parallelization using MPI+X, where X can be OpenMP for multicore machines, or CPU/GPU systems~\cite{zhang2019amrex}. The source code for IAMR and all testing cases used in this work can be accessed at~\url{https://github.com/ruohai0925/IAMR/tree/development}. All postprocessing scripts and profiling data are also available at~\url{https://github.com/Echo-Lau/IAMR_Tutorial_Profiling_Results}.

\section{Parameters related to BSAMR} \label{S:3}

This section introduces five important parameters considered in this work while conducting the simulations with BSAMR.

\begin{list}{$\circ$}{}

\item ${Max\_level}$: In BSAMR, the ${Max\_level}$ refers to maximum level of refinement. For the single-level simulation, we have ${Max\_level}=0$. For all of the multi-level simulations in Section~\ref{S:5}, ${Max\_level}$ is set to be 1 or 2, which is common in many physical simulations using BSAMR~\cite{zeng2021subcycling,bhalla2013unified}. To make the meshes finer level enough to capture the flow physics accurately, one can use either more levels of refinement (i.e., a large ${Max\_level}$) or directly set the number of cells on the coarsest level to be large. In this work, we choose the second option.    

\item ${Max\_grid\_size}$: The load balancing algorithm divides the domain in each direction so that each grid/patch is no longer than ${Max\_grid\_size}$ in that direction. Note that ${Max\_grid\_size}$ is just an upper bound. If the number of cells in $x$-direction is 48 and the ${Max\_grid\_size}$ is 32, we will typically have one grid/patch of length 32 and one of length 16. It is seen that small ${Max\_grid\_size}$ leads to large grid/patches, which then need to be distributed to multiple CPUs for parallel computation~\cite{zhang2019amrex}.

\item ${Regrid\_interval}$: The ${Regrid\_interval}$ represents how often to refine or de-refine the meshes (in terms of steps on the coarsest level) during the simulation. Large ${Regrid\_interval}$ refers to less frequent refinement, which means the static meshes can stay longer. Small ${Regrid\_interval}$ means more frequent changes of meshes, which brings additional work for communications across different processors~\cite{zhang2020amrex}.

\item ${Cycling}$: As shown in Section~\ref{S:23}, either the subcycling or the non-subcycling method can be used for simulations with BSAMR.

\item ${Skip\_level\_projection}$: Since the time advancement is done level by level on the multi-level grid, the ${Skip\_level\_projection}$ is applied to check whether the pressure projection steps on coarse levels can be omitted without bringing side effects of the whole simulation. Let us say we have two levels, i.e., level 0 and level 1. If ${Skip\_level\_projection=0}$, then the level projection is applied to both level 0 and level 1. If ${Skip\_level\_projection=1}$, then level projection is applied only to level 1. 

\end{list} 

\section{Testing Cases} \label{S:4}

We tested various two-dimensional and three-dimensional cases (Table~\ref{tab:testcases}), categorizing each case into thirty-two distinct testing scenarios based on different parameters.~Taking the 2D lid-driven case as an example, Table~\ref{tab:test32} displays the thirty-two combinations of parameters for this case, while each combination corresponds to the different parameter values listed in Session~\ref{S:3}.~To compare computational efficiency, we repeatedly ran each tests several times and recorded the average running time.~In addition, for a specific case, we conducted tests using different CPUs and GPUs (Table~\ref{tab:computer_architecture}) to investigate the impact of software parameters on running efficiency under different computer hardware.

\begin{table}[H]
  \centering
  \caption{Testing cases in this work (DSL: Double Shear Layer; FPC: Flow Past Cylinder; LDC: Lid Driven Cavity; TGV: Taylor Green Vortex).}
  {
  \fontsize{8}{6}\selectfont 
  \begin{tabular}{c|c|c|c|c|c|c|c} 
    \toprule 
    Dimension & Bubble &  Convected Vortex & DSL & FPC & LDC & RT & TGV \\
    \midrule 
    2D & Yes & Yes & Yes & Yes & Yes & Yes & Yes \\
    3D & No & Yes & No & Yes & Yes & No & Yes \\
    \bottomrule 
  \end{tabular}
  }
  
  \label{tab:testcases}
\end{table}

\begin{table}[H]
  \centering
  \caption{Thirty-two combinations of parameters for each case}
  {
  \fontsize{9}{7}\selectfont 
  \begin{tabular}{c|c|c|c|c|c} 
    \toprule 
    Test No. & Skip level projection & Cycling & Max level & Max grid size & Regrid interval \\
    \midrule 
    1 & 0 & Auto & 1 & 8 & 4 \\
    2 & 0 & Auto & 1 & 8 & 8 \\
    3 & 0 & Auto & 2 & 8 & 4 \\
    4 & 0 & Auto & 2 & 8 & 8 \\
    5 & 0 & Auto & 1 & 16 & 4 \\
    6 & 0 & Auto & 1 & 16 & 8 \\
    7 & 0 & Auto & 2 & 16 & 4 \\
    8 & 0 & Auto & 2 & 16 & 8 \\
    9 & 0 & None & 1 & 8 & 4 \\
    10 & 0 & None & 1 & 8 & 8 \\
    11 & 0 & None & 2 & 8 & 4 \\
    12 & 0 & None & 2 & 8 & 8 \\
    13 & 0 & None & 1 & 16 & 4 \\
    14 & 0 & None & 1 & 16 & 8 \\
    15 & 0 & None & 2 & 16 & 4 \\
    16 & 0 & None & 2 & 16 & 8 \\
    17 & 1 & Auto & 1 & 8 & 4 \\
    18 & 1 & Auto & 1 & 8 & 8 \\
    19 & 1 & Auto & 2 & 8 & 4 \\
    20 & 1 & Auto & 2 & 8 & 8 \\
    21 & 1 & Auto & 1 & 16 & 4 \\
    22 & 1 & Auto & 1 & 16 & 8 \\
    23 & 1 & Auto & 2 & 16 & 4 \\
    24 & 1 & Auto & 2 & 16 & 8 \\
    25 & 1 & None & 1 & 8 & 4 \\
    26 & 1 & None & 1 & 8 & 8 \\
    27 & 1 & None & 2 & 8 & 4 \\
    28 & 1 & None & 2 & 8 & 8 \\
    29 & 1 & None & 1 & 16 & 4 \\
    30 & 1 & None & 1 & 16 & 8 \\
    31 & 1 & None & 2 & 16 & 4 \\
    32 & 1 & None & 2 & 16 & 8 \\
    \bottomrule 
  \end{tabular}
  }
  \label{tab:test32}
\end{table}

\begin{table}[H]
    \centering
    \caption{Computational hardware}
    \fontsize{6}{6}\selectfont 
        \begin{tabular}{c|c|c|c|c}
        \hline
        Computational hardware  &  AMD Ryzen 5 5500U &  AMD Ryzen 9 7845HX &  Tesla V100-SXM2-16GB &  RTX 4060 \\
        \hline
        Detail & CPU / 6 cores / 16 GB & CPU / 12 cores / 16 GB    & GPU / 80 cores / 16 GB & GPU / 24 cores / 8 GB \\
        \hline
        \end{tabular}
        \label{tab:computer_architecture}
\end{table}

\section{Results} \label{S:5}


Due to the limited space in the paper, we have selected results of representative cases to explain the influence of specific software-based parameters and provide the insights behind them. We have uploaded the scripts used for processing the codes and all post-processed data to GitHub, such that interested readers can review and reproduce the work presented in this paper. The GitHub repository of post-processing data can be accessed through~\url{https://github.com/Echo-Lau/IAMR_Tutorial_Profiling_Results}.

\subsection{CPU and GPU performance} \label{S:51}

\begin{figure}[H]
    \centering
    \includegraphics[width=1.0\linewidth]{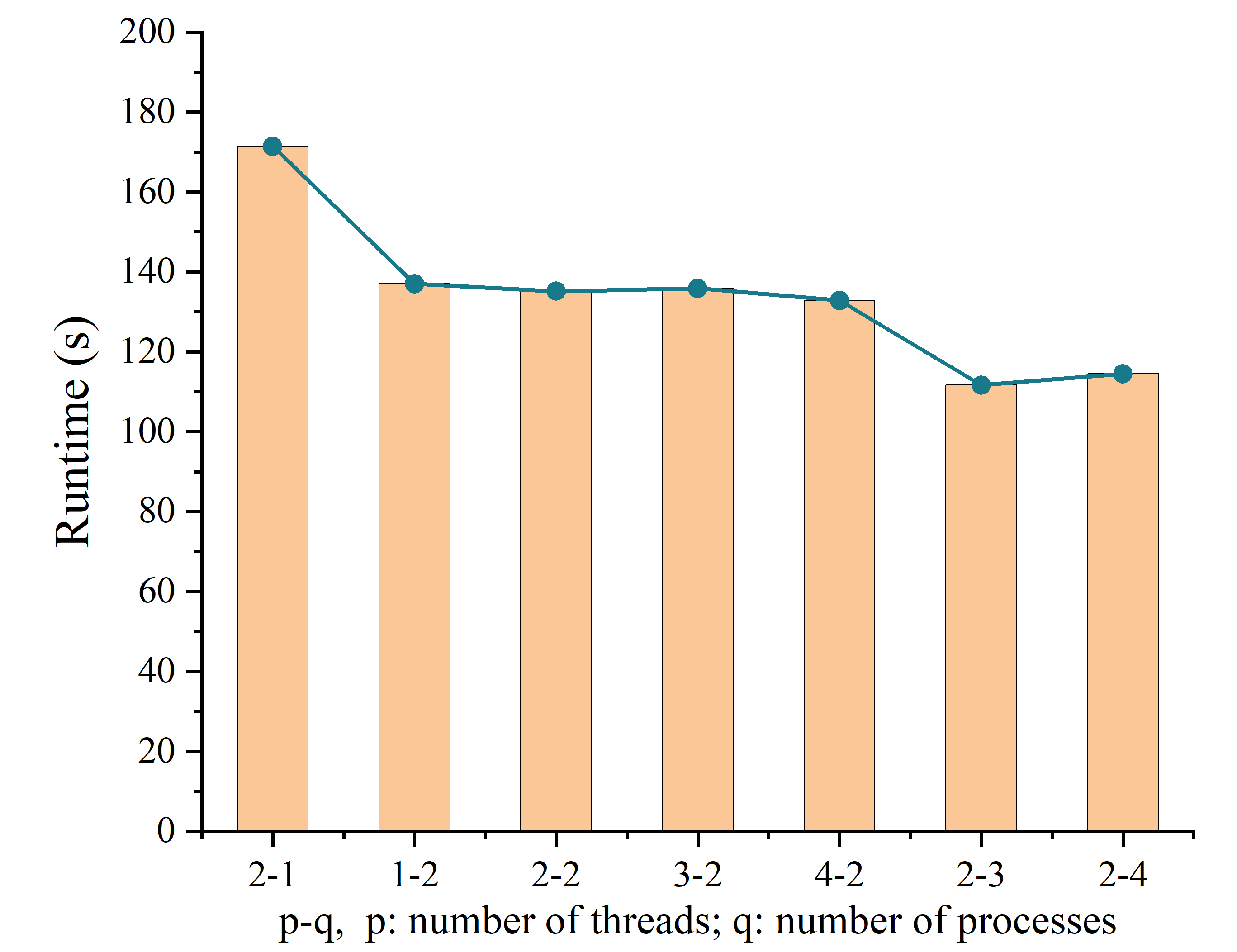}
    \caption{Impact of multithread/multicore CPUs on the runtime of lid-driven cavity case}
    \label{fig:cpu1}
\end{figure}

In the field of computational fluid dynamics (CFD), CPU parallel computing plays an important role. For tests on the CPU, we used an AMD R5 5500 model type, which features 6 physical cores and hyper-threading technology. As it utilizes shared memory, it supports testing the efficiency of IAMR code execution within a mixed OpenMP and MPI programming environment~\cite{zhang2019amrex}. Figure~\ref{fig:cpu1} presents the runtime of a two-dimensional lid-driven case under different thread and process combinations, where the x-axis labeled p-q represents the use of p threads and q processes. We observed that an increase in MPI processes significantly reduces computation time, whereas an increase in the number of OpenMP threads results in no significant change in runtime. This indicates that within the framework of the IAMR software, OpenMP has a negligible impact on computational efficiency, whereas increasing the number of MPI processes can significantly enhance it. At the same time, as illustrated by the chart for 2-4 in Figure~\ref{fig:cpu1}, where increasing the number of MPI processes beyond a certain level, even without exceeding the maximum number of hyperthreading cores, results in increased runtime. Therefore, using more MPI processes is not always better, especially when the number of grid cells is low. It is important to select an appropriate number of processes for a specific case.

\begin{figure}[H]
    \centering
    \begin{subfigure}{0.5\textwidth}
        \centering
        \includegraphics[width=\textwidth]{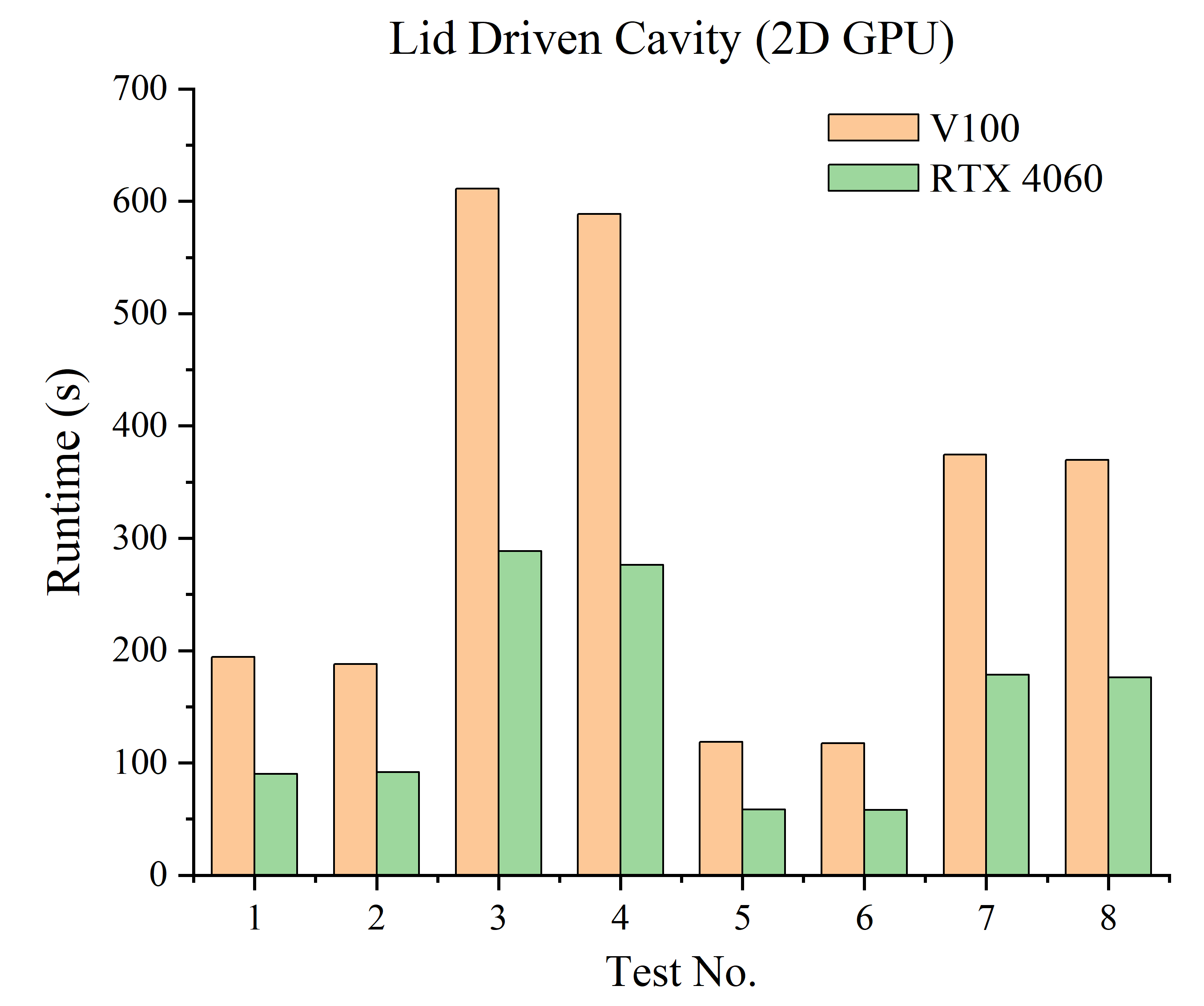}
    \end{subfigure}\hfill
    \begin{subfigure}{0.5\textwidth}
        \centering
        \includegraphics[width=\textwidth]{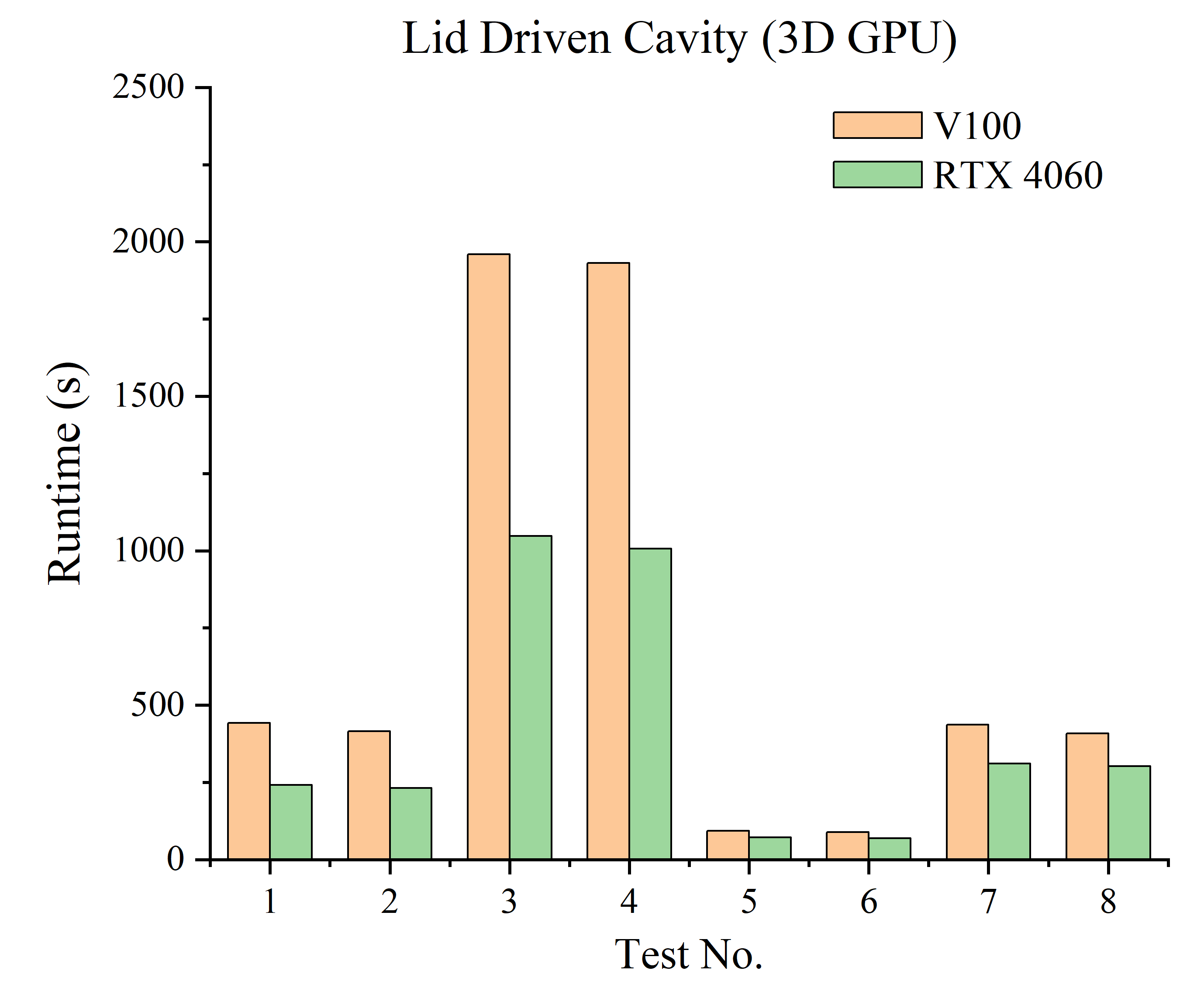}       
    \end{subfigure}\hfill
    \caption{Impact of different GPUs on the runtime of lid-driven cavity case}
    \label{fig:gpu1}
\end{figure}

In terms of GPU testing, we utilized the Tesla V100-SXM2 and the RTX 4060 to assess the computational efficiency. We found that the computation time required on the V100 was longer for most cases, particularly noticeable in two-dimensional scenarios. For three-dimensional cases, the difference in runtime between the two GPUs decreases (Figure~\ref{fig:gpu1}). This could be attributed to the RTX 4060 employing NVIDIA's newer architecture (such as Ampere or later) and faster memory technology (such as GDDR6 or GDDR6X). Also, compute-intensive tasks might fully leverage the high-frequency CUDA cores of the RTX 4060 rather than relying on the larger quantity of lower-frequency CUDA cores provided by the V100, resulting in better performance from the RTX 4060.

\begin{figure}[H]
    \centering
    \includegraphics[width=1.0\linewidth]{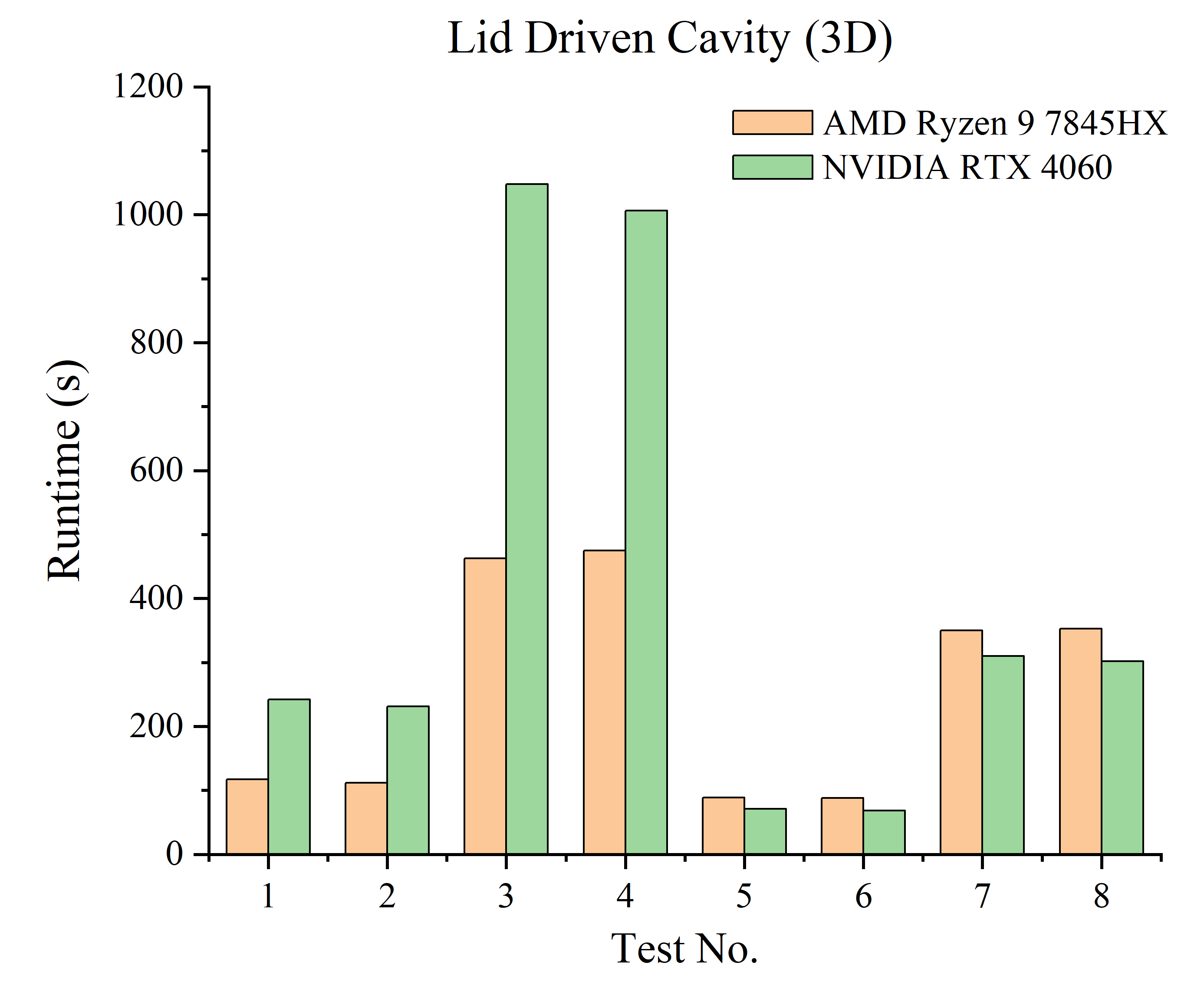}
    \caption{Comparison of runtime on the CPU and GPU for the 3D lid-driven cavity case}
    \label{fig:cpugpu}
\end{figure}

\begin{figure}[H]
    \centering  
    \begin{subfigure}{0.5\textwidth}
        \centering
        \includegraphics[width=\textwidth]{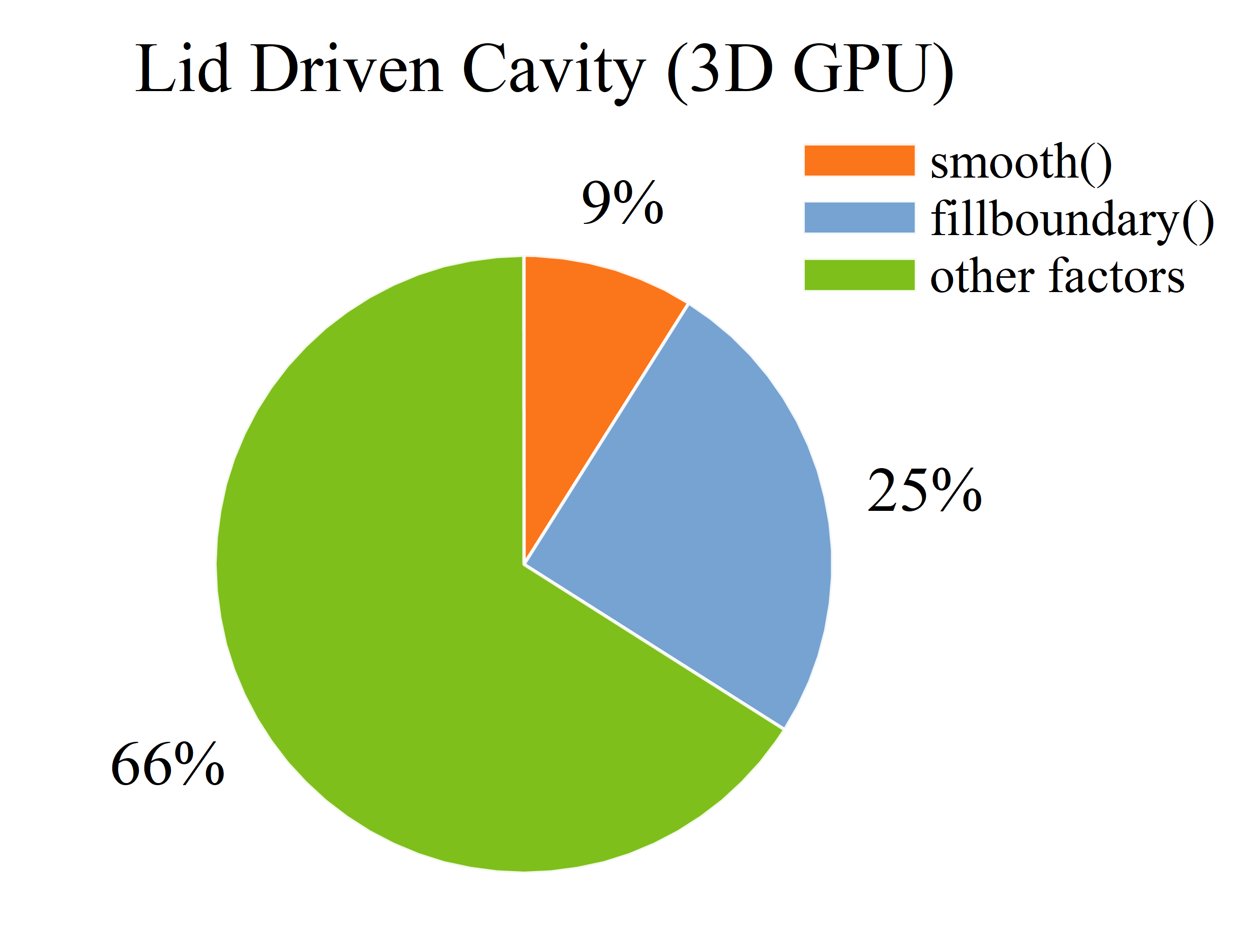}
        \caption{Max level = 8}
        \label{fig:function_a}
    \end{subfigure}\hfill
    \begin{subfigure}{0.5\textwidth}
        \centering
        \includegraphics[width=\textwidth]{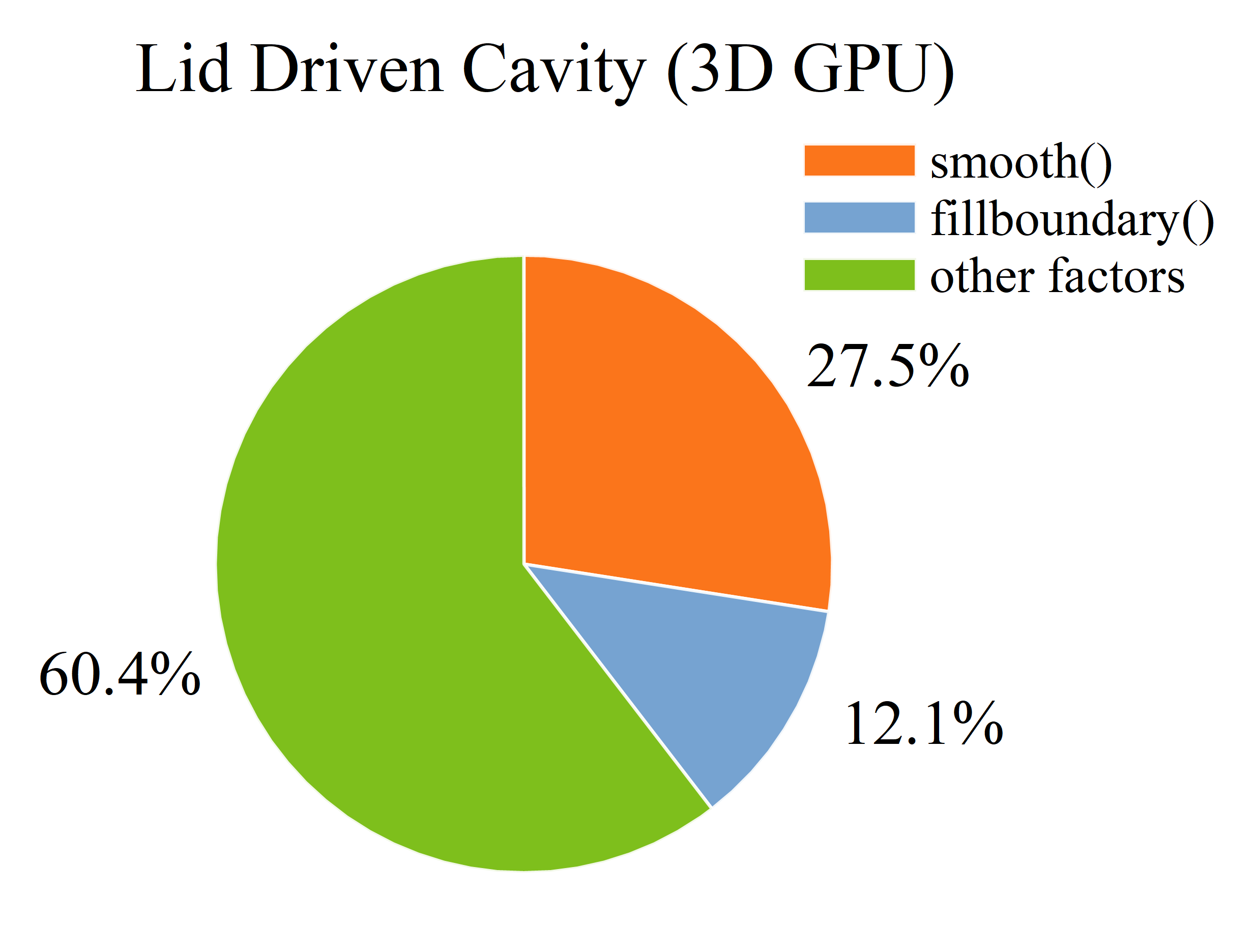}
        \caption{Max level = 16}
        \label{fig:function_b}
    \end{subfigure}
    \caption{Percentage of function call time on GPUs for the 3D lid-driven cavity case}
    \label{fig:function}
\end{figure}

We also found that in most cases, a CPU configuration with 4 processes and 2 threads runs faster than a GPU using the IAMR framework. Yet, for some three-dimensional cases with a large ${Max\_grid\_size}$ of 16, GPU execution can surpass that of the CPU. The bar charts in Figure~\ref{fig:cpugpu} with the x-axis labeled 5, 6, 7, and 8 provide evidence for this observation. The reason is that the number of grid blocks decreases for larger ${Max\_grid\_size}$ values, which then reduces communication overhead and allows the computation to dominate. And thus the operational advantages of GPUs are highlighted. To further confirm this point, we conducted a detailed analysis of specific function's call times as a percentage of total runtime in some examples. As shown in Figure~\ref{fig:function}, the function fillboundary() represents communication-related tasks, and the function smooth() pertains to computation-related tasks. We observed that for smaller cases with fewer grid numbers, communication time dominates. However, as cases transition to three-dimensional with an increased number of grids and a larger ${Max\_grid\_size}$, the GPU's runtime becomes less than the CPU's.

\subsection{Max\_level} \label{S:52}
\begin{figure}[H]
\centering
    \begin{subfigure}{0.5\textwidth}
      \centering
      \includegraphics[width=\textwidth]{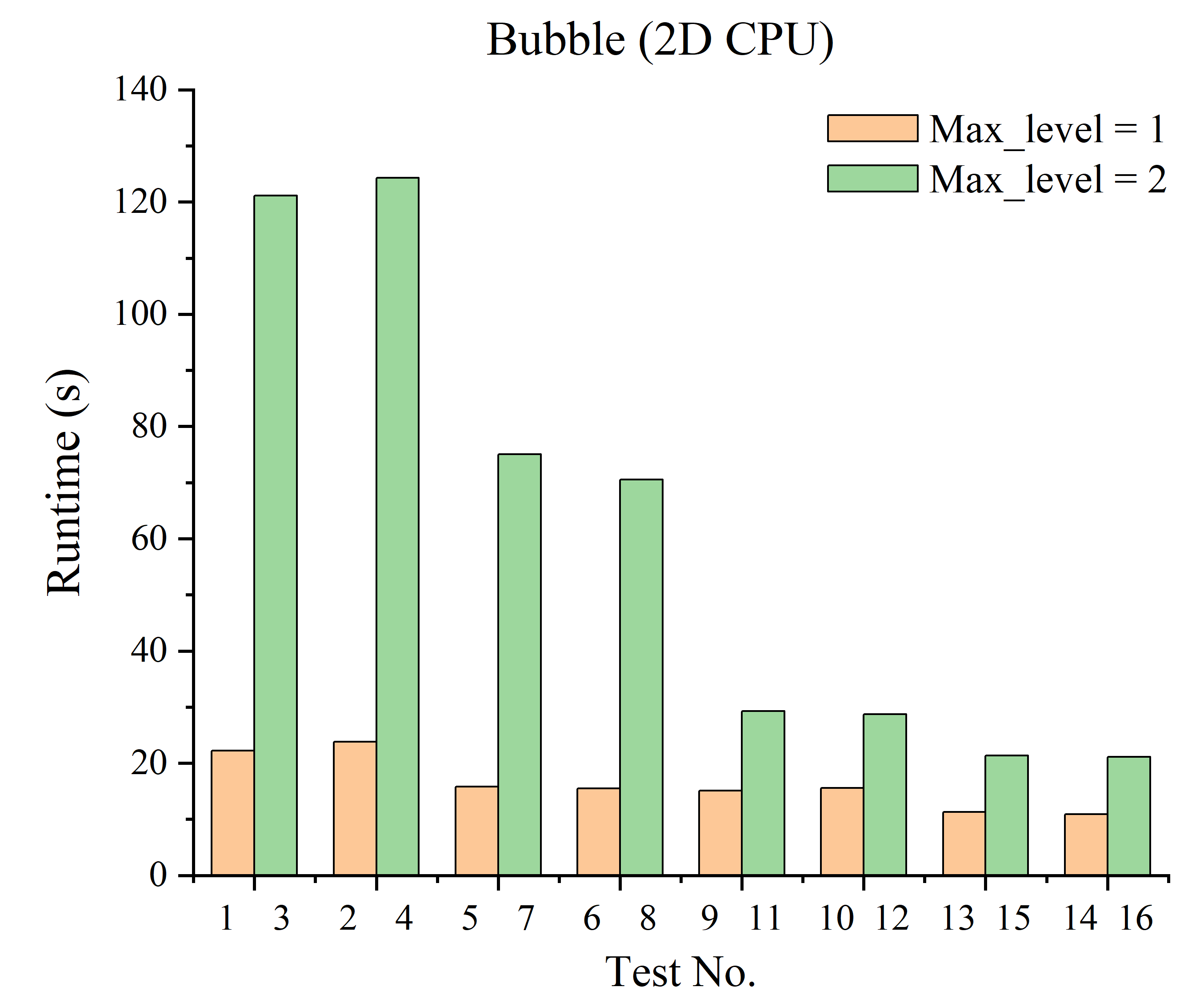}
      \caption{}
      \label{fig:maxlevel_a}
    \end{subfigure}\hfill
    \begin{subfigure}{0.5\textwidth}
      \centering
      \includegraphics[width=\textwidth]{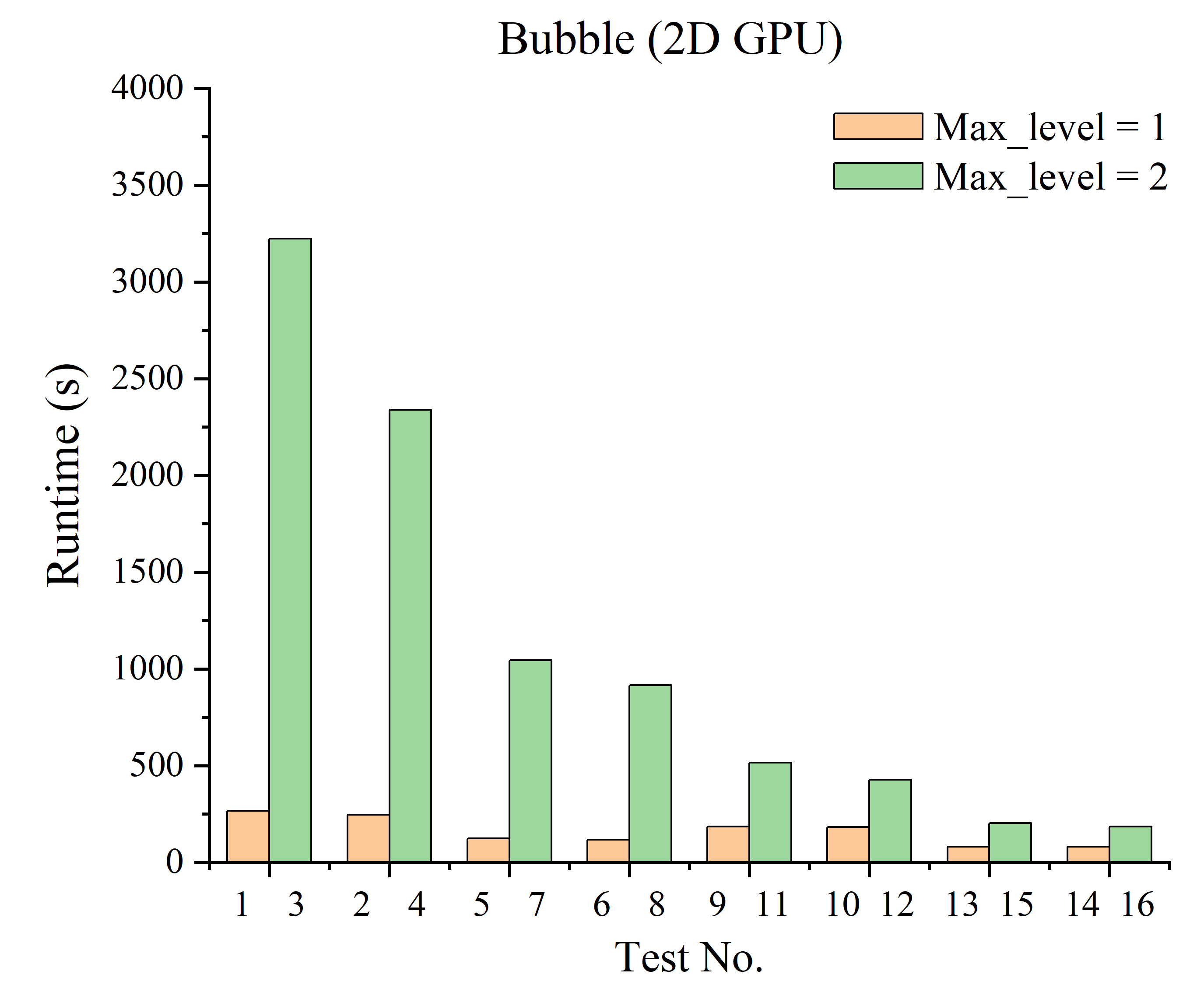}
      \caption{}
      \label{fig:maxlevel_b}
    \end{subfigure}
\caption{Running time for various cases with different ${Max\_level}$}
\label{fig:maxlevel}
\end{figure}

The influence of the ${Max\_level}$ parameter on runtime across different cases is examined in Figure~\ref{fig:maxlevel}. Taking panel (a) as an example, its title "Bubble (2D CPU)" indicates a two-dimensional Bubble case running on a CPU. The x-axis, labeled "Test No.", corresponds to different testing scenarios (which can be cross-referenced with Table~\ref{tab:test32} in Section~\ref{S:4}), and the y-axis represents the runtime. The setup for the other panels is similar to panel (a). Regarding the ${Max\_level}$ parameter, we observed that across all cases, regardless of whether they run on a CPU or GPU, a ${Max\_level}$ of 1 always resulted in shorter runtime than a ${Max\_level}$ of 2, due to the former having fewer grid cells.

\subsection{Max\_grid\_size} \label{S:53}
\begin{figure}[H]
\centering
\begin{subfigure}{0.5\textwidth}
  \centering
  \includegraphics[width=\textwidth]{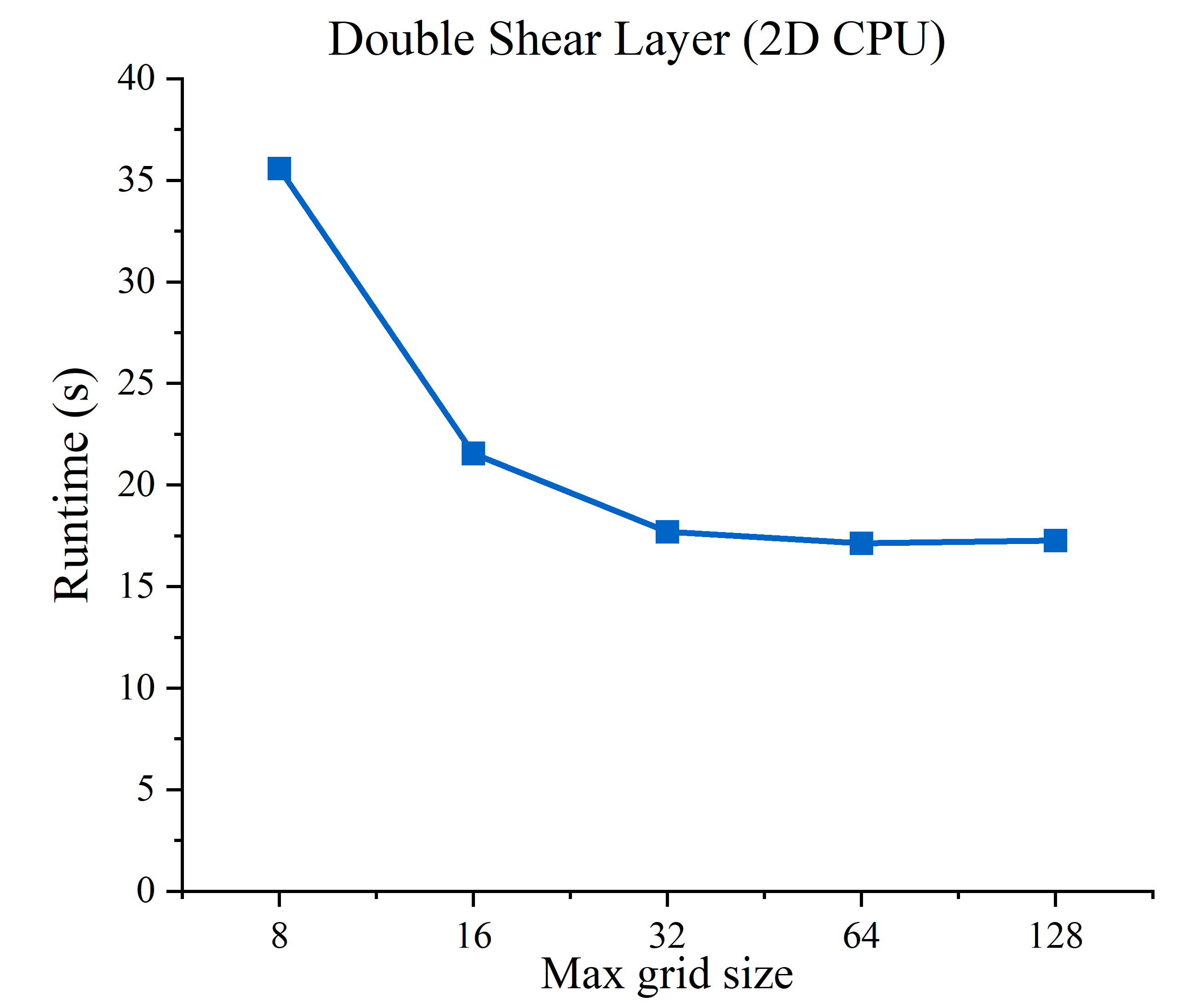}
  \caption{}
  \label{fig:maxgridsize_a}
\end{subfigure}\hfill
\begin{subfigure}{0.5\textwidth}
  \centering
  \includegraphics[width=\textwidth]{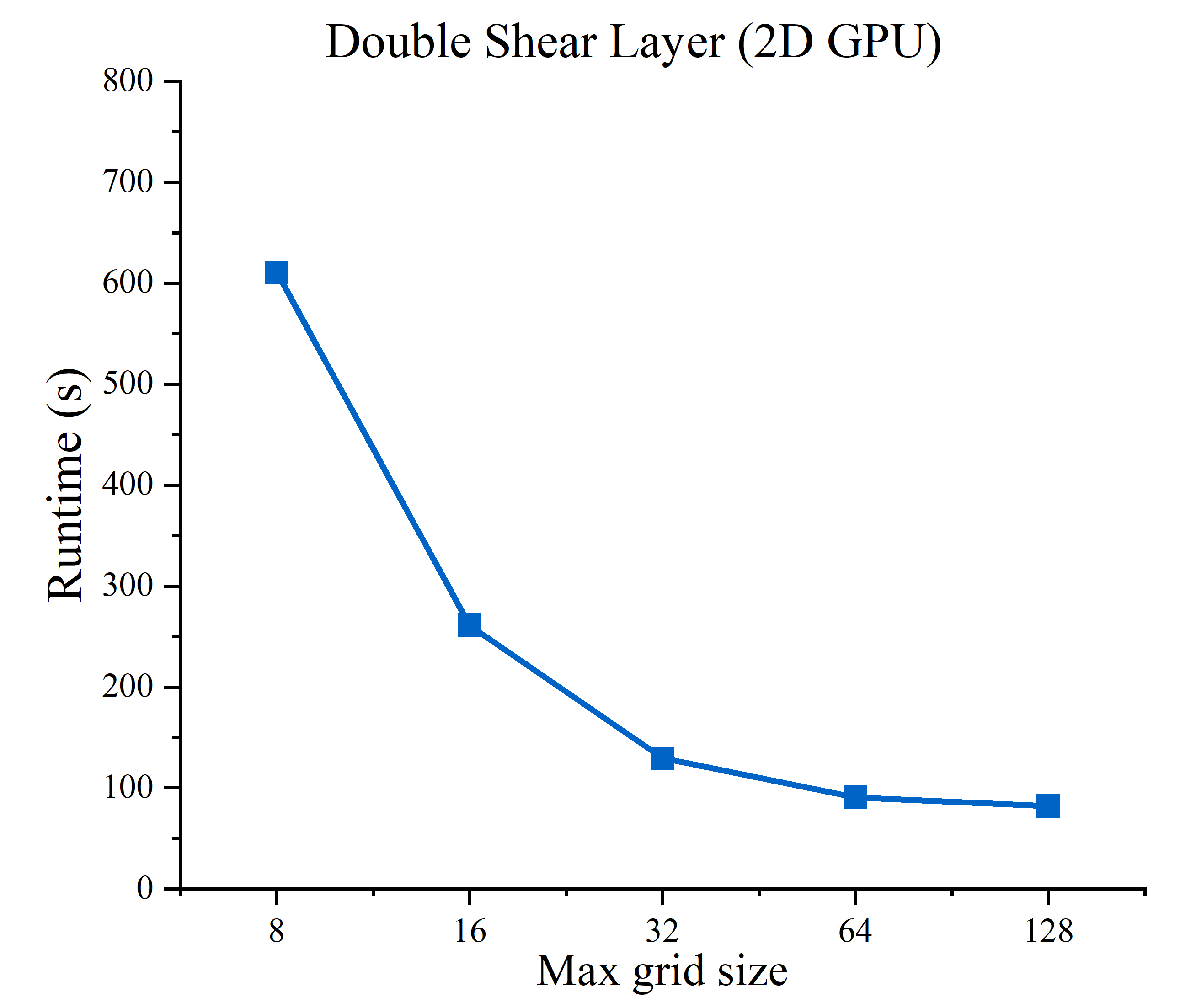}
  \caption{}
  \label{fig:maxgridsize_b}
\end{subfigure}

\caption{Running time for various cases with different ${Max\_grid\_size}$}
\label{fig:maxgridsize}
\end{figure}

Taking the two-dimensional Double Shear Layer (DSL) case as an example, we kept other parameters constant and compared the runtime across different tests with ${Max\_grid\_size}$ values of 8, 16, 32, 64, and 128. The results in Figure~\ref{fig:maxgridsize} indicate that as the ${Max\_grid\_size}$ increases, the runtime gradually decreases. This improvement in running efficiency primarily stems from the reduced communication time between different grid boxes. Since the code running on a GPU is more sensitive to the overhead caused by communication, this pattern is more pronounced when running on a GPU (as seen in panel (b)). Additionally, when the ${Max\_grid\_size}$ reaches a certain threshold, the reduction in runtime becomes marginal. This is mainly because, at this point, the ${Max\_grid\_size}$ exceeds the box size at the root level (i.e., level 0), making the communication overhead nearly negligible. 

\subsection{Cycling} \label{S:54}

To investigate the effect of the cycling method on running efficiency, we further subdivided the tests for a particular case into two different scenarios. One scenario involves running all thirty-two tests to the maximum step, e.g., $max\_step = 200$. The other scenario has all thirty-two tests run for the same simulation duration, with these tests running for 1.0 seconds, for example. From the cases shown in the following Figure~\ref{fig:cycling}, we observed that for cases completing the maximum number of steps ($max\_step$), regardless of whether they are two-dimensional or three-dimensional, or whether they run on a CPU or GPU, the running time with $cycling=Auto$ (i.e., using the subcycling method) is always longer than with $cycling=None$ (i.e., using the non-subcycling method). This is expected because, with the subcycling method, finer levels always run more sub-steps, thus increasing the total running time. For cases completing the stop time, the running time with $cycling=Auto$ is shorter than with $cycling=None$ because the subcycling method allows different levels to use different time steps, enabling coarse levels to use larger time steps for fast advancement. The insights here suggest that for a steady-state simulation, assuming the simulation duration needed to reach steady-state is known, one could expediently advance using the subcycling method, provided the CFL stability condition is met. Conversely, if the goal is merely to reach a fixed step count to check the test results, a non-subcycling method could be utilized.

\begin{figure}[H]
\centering
\begin{subfigure}{0.5\textwidth}
  \centering
  \includegraphics[width=\textwidth]{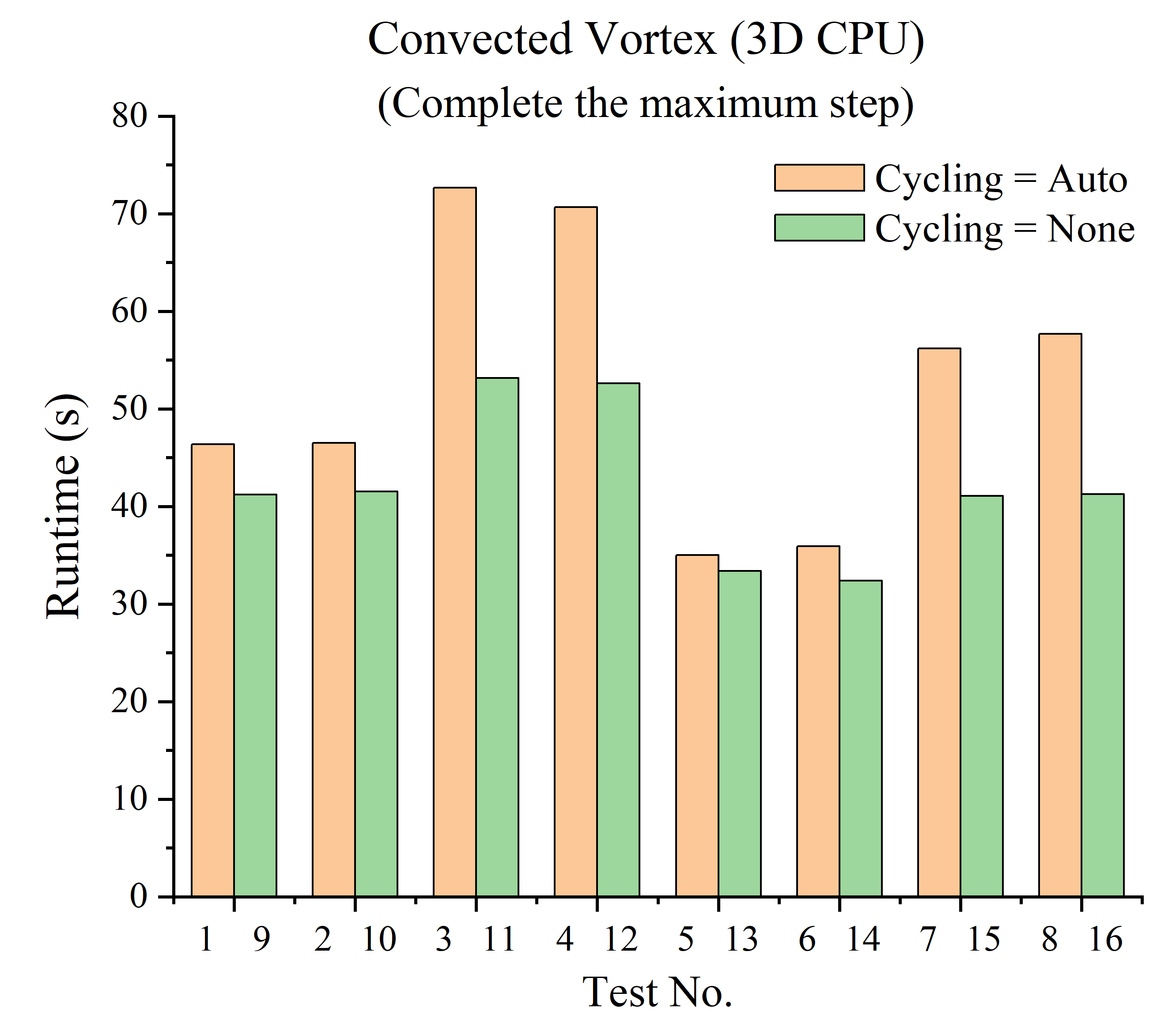}
  \caption{}
  \label{fig:cycling_a}
\end{subfigure}\hfill
\begin{subfigure}{0.5\textwidth}
  \centering
  \includegraphics[width=\textwidth]{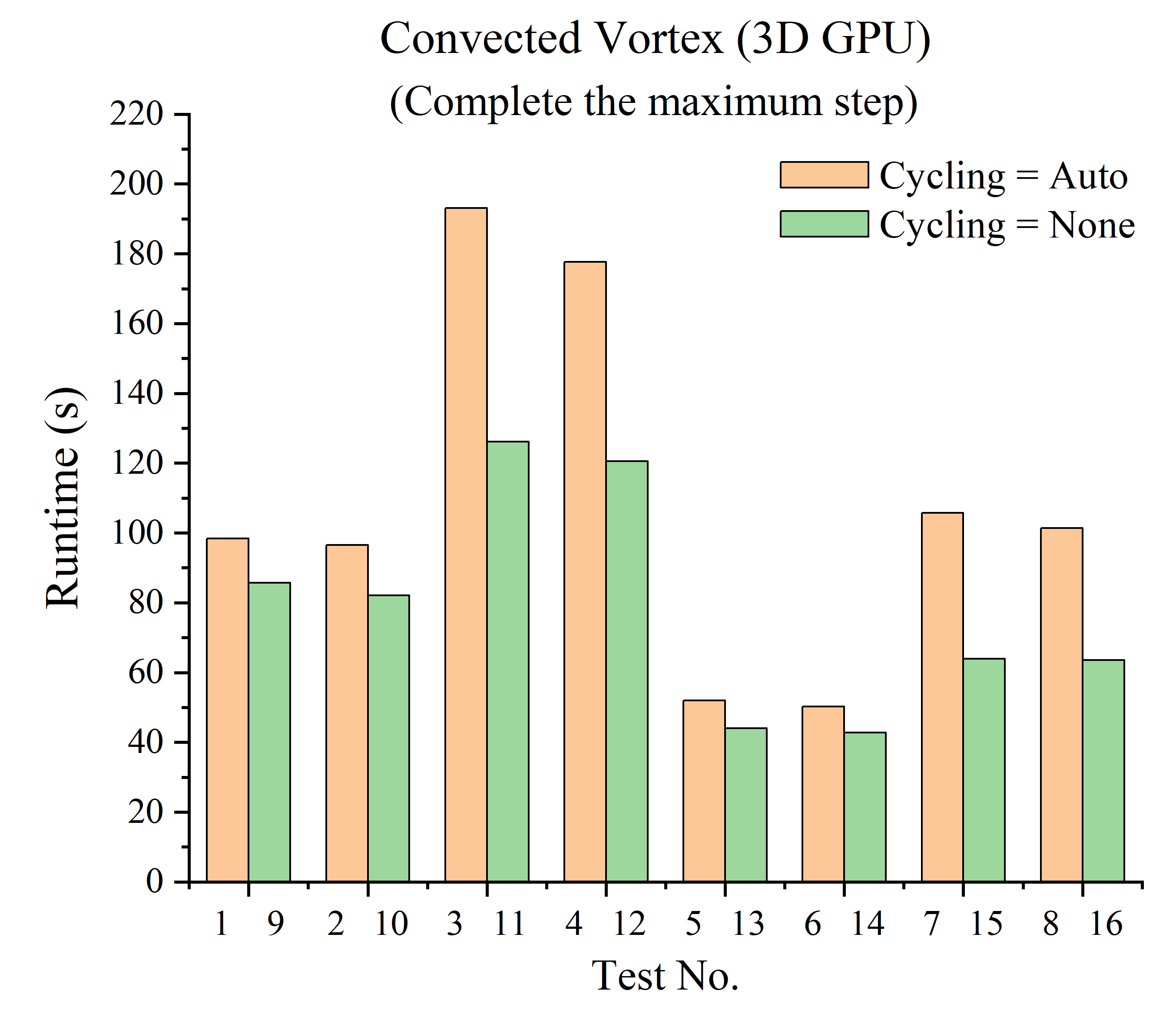}
  \caption{}
  \label{fig:cycling_b}
\end{subfigure}
\begin{subfigure}{0.5\textwidth}
  \centering
  \includegraphics[width=\textwidth]{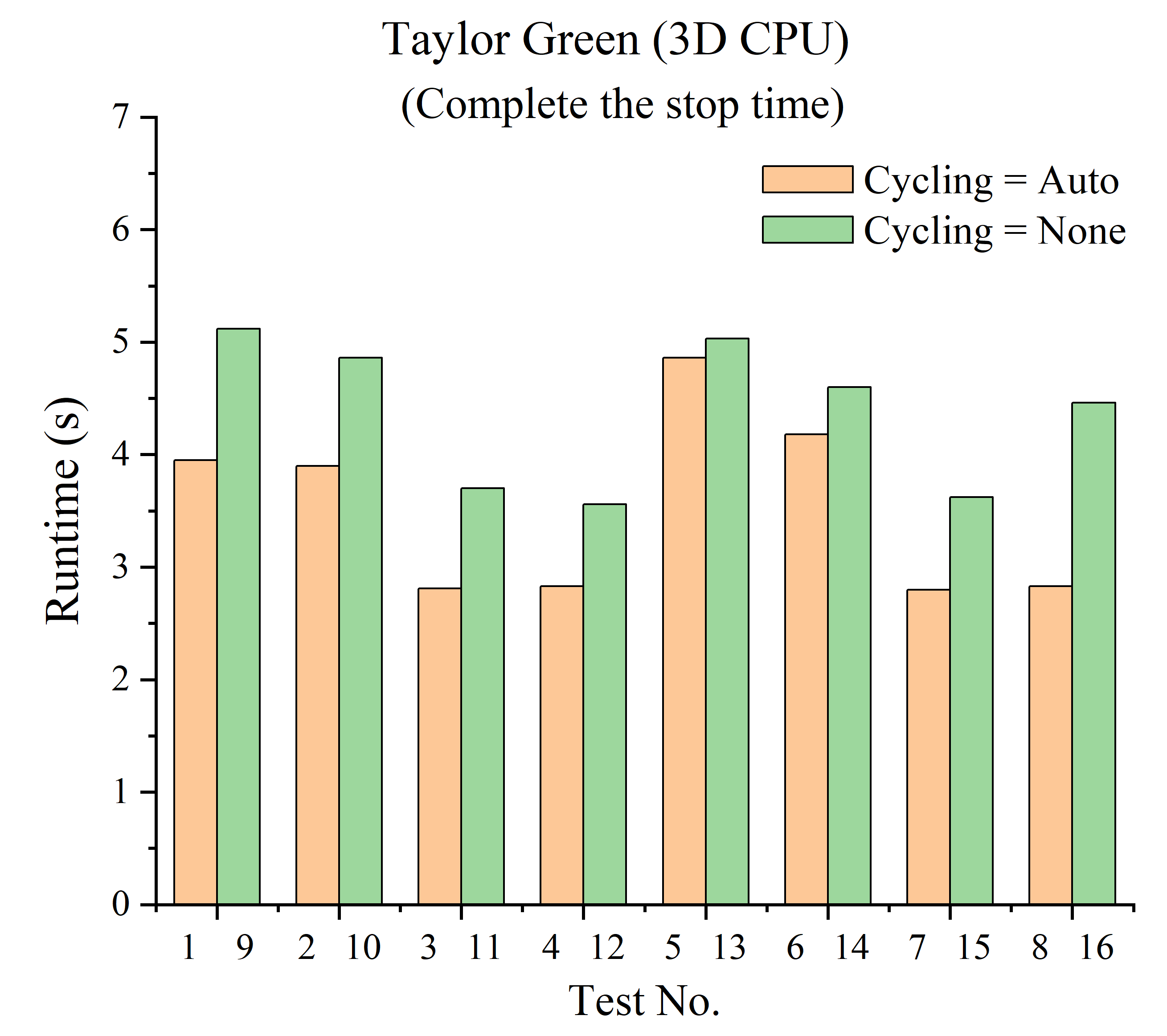}
  \caption{}
  \label{fig:cycling_c}
\end{subfigure}\hfill
\begin{subfigure}{0.5\textwidth}
  \centering
  \includegraphics[width=\textwidth]{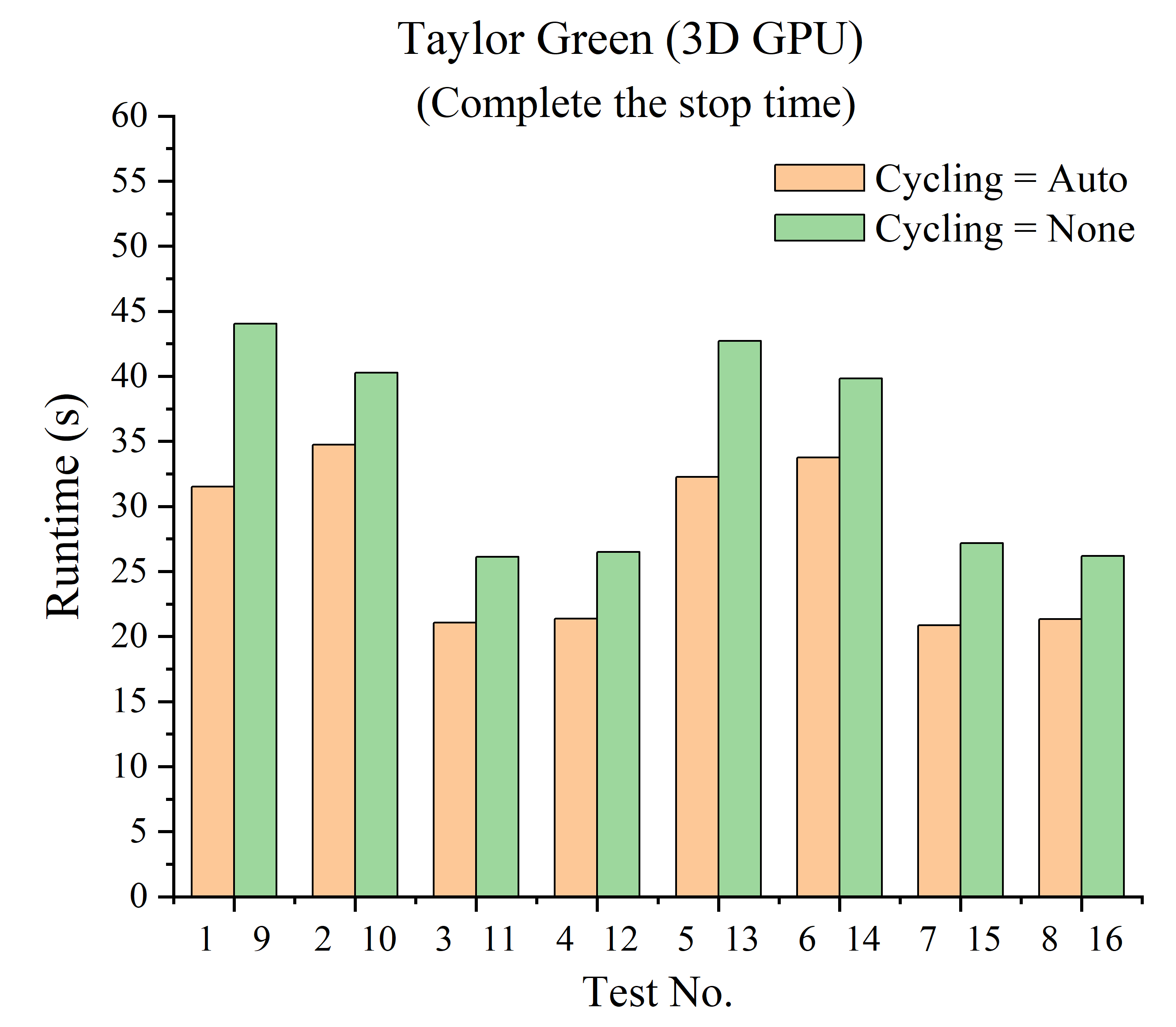}
  \caption{}
  \label{fig:cycling_d}
\end{subfigure}
\caption{Running time for various cases with different ${Cycling}$}
\label{fig:cycling}
\end{figure}

\subsection{Regrid\_interval} \label{S:55}
\begin{figure}[H]
\centering
\begin{subfigure}{0.5\textwidth}
  \centering
  \includegraphics[width=\textwidth]{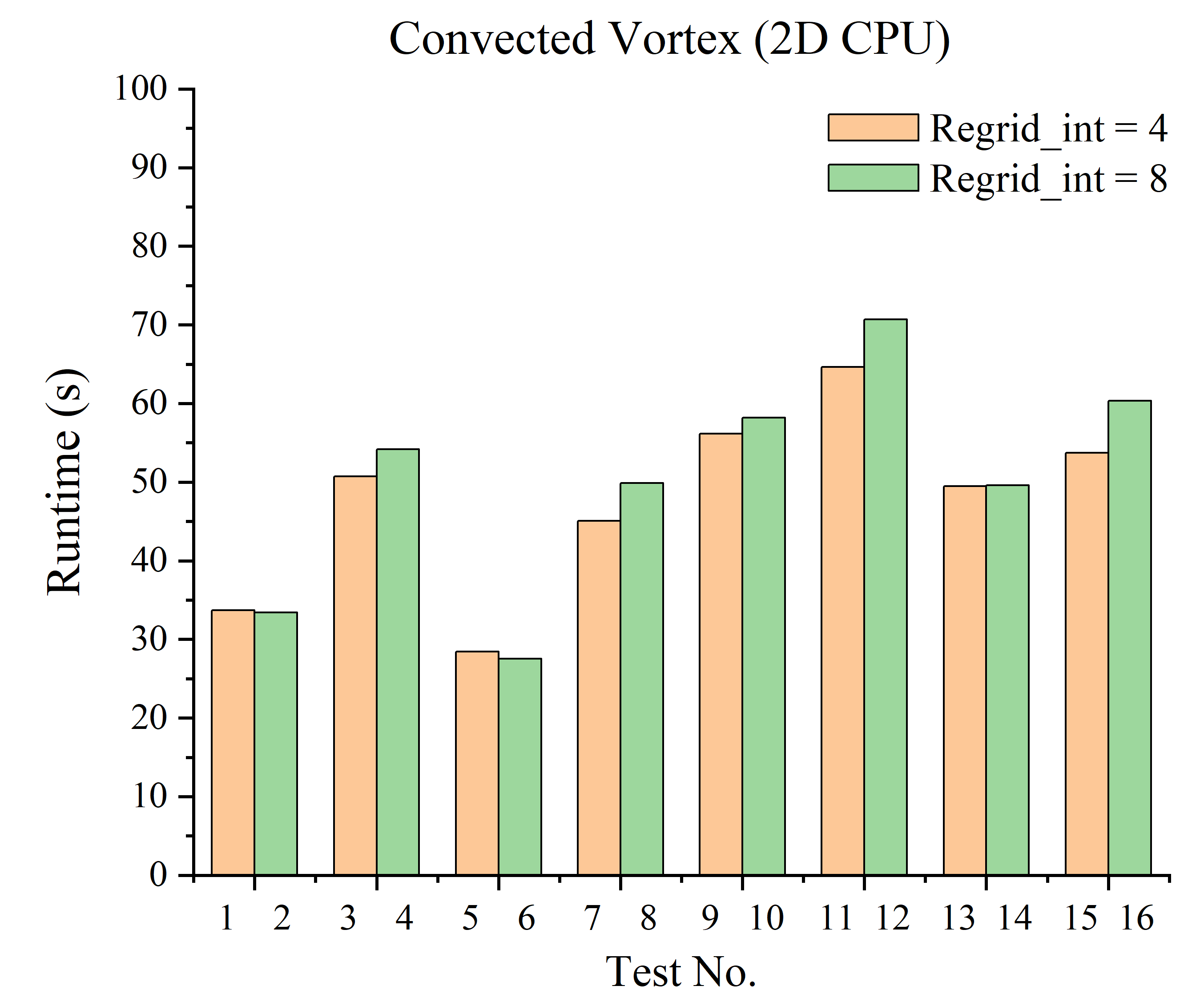}
  \caption{}
  \label{fig:regrid_int_a}
\end{subfigure}\hfill
\begin{subfigure}{0.5\textwidth}
  \centering
  \includegraphics[width=\textwidth]{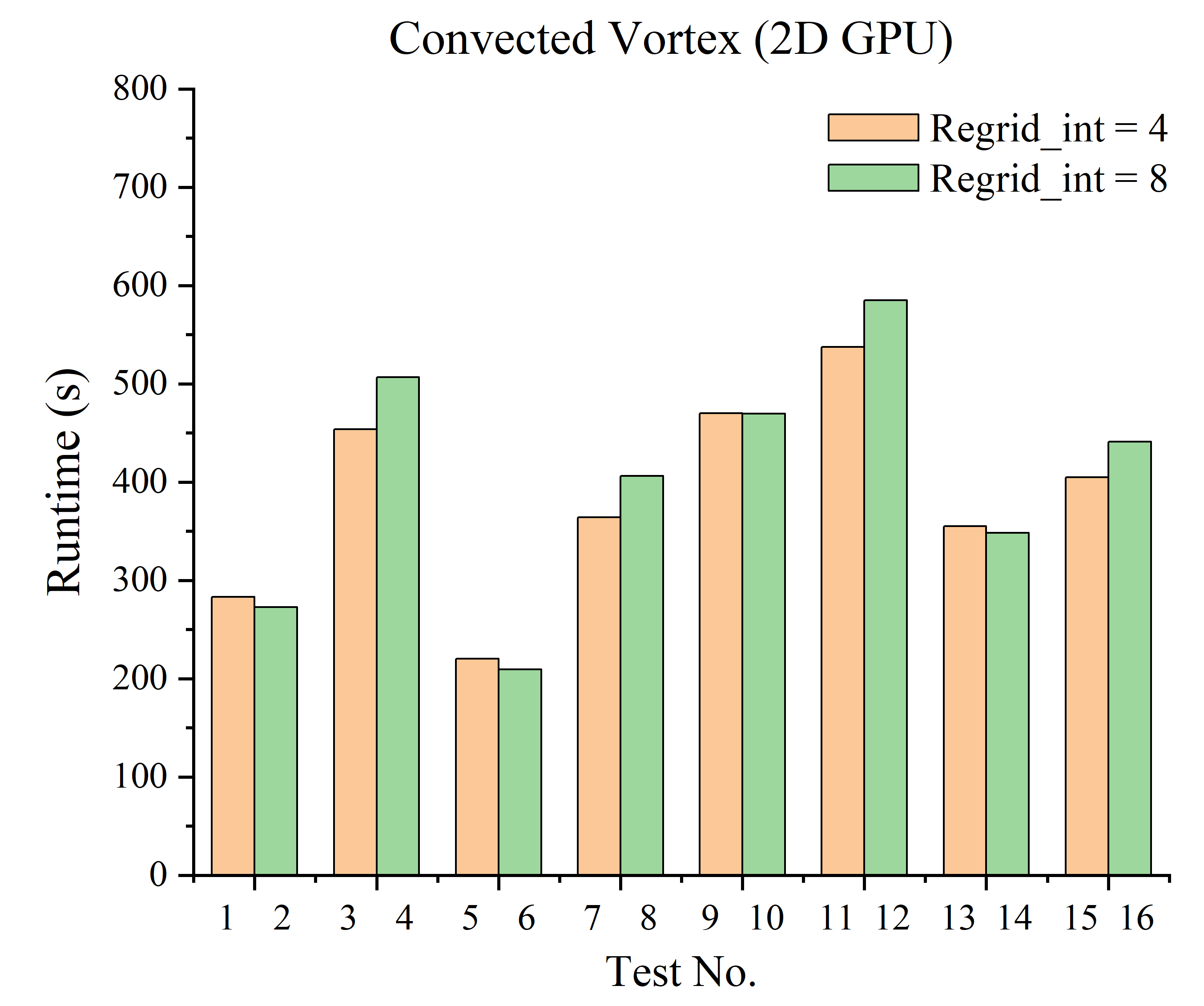}
  \caption{}
  \label{fig:regrid_int_b}
\end{subfigure}

\begin{subfigure}{0.5\textwidth}
  \centering
  \includegraphics[width=\textwidth]{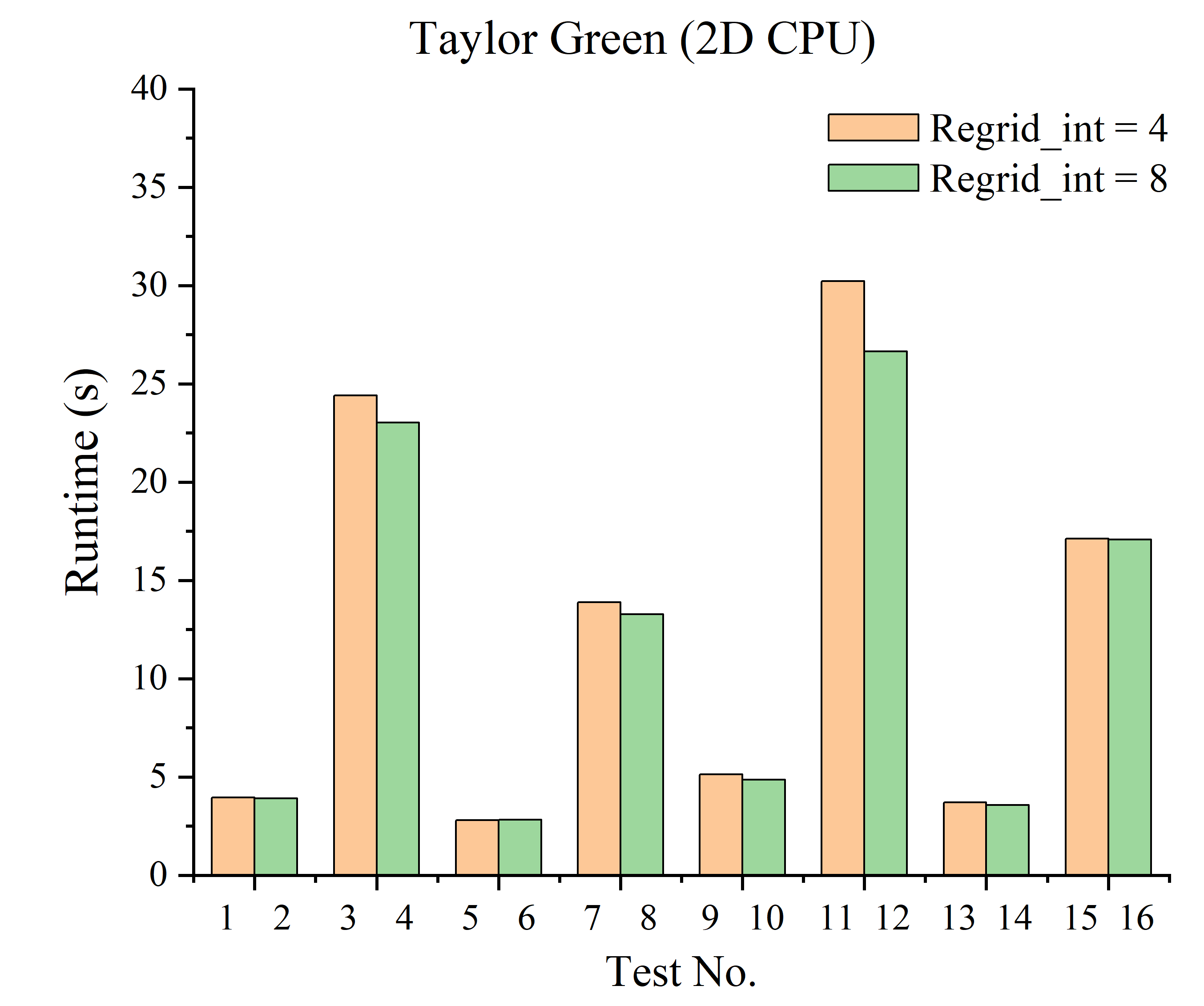}
  \caption{}
  \label{fig:regrid_int_c}
\end{subfigure}\hfill
\begin{subfigure}{0.5\textwidth}
  \centering
  \includegraphics[width=\textwidth]{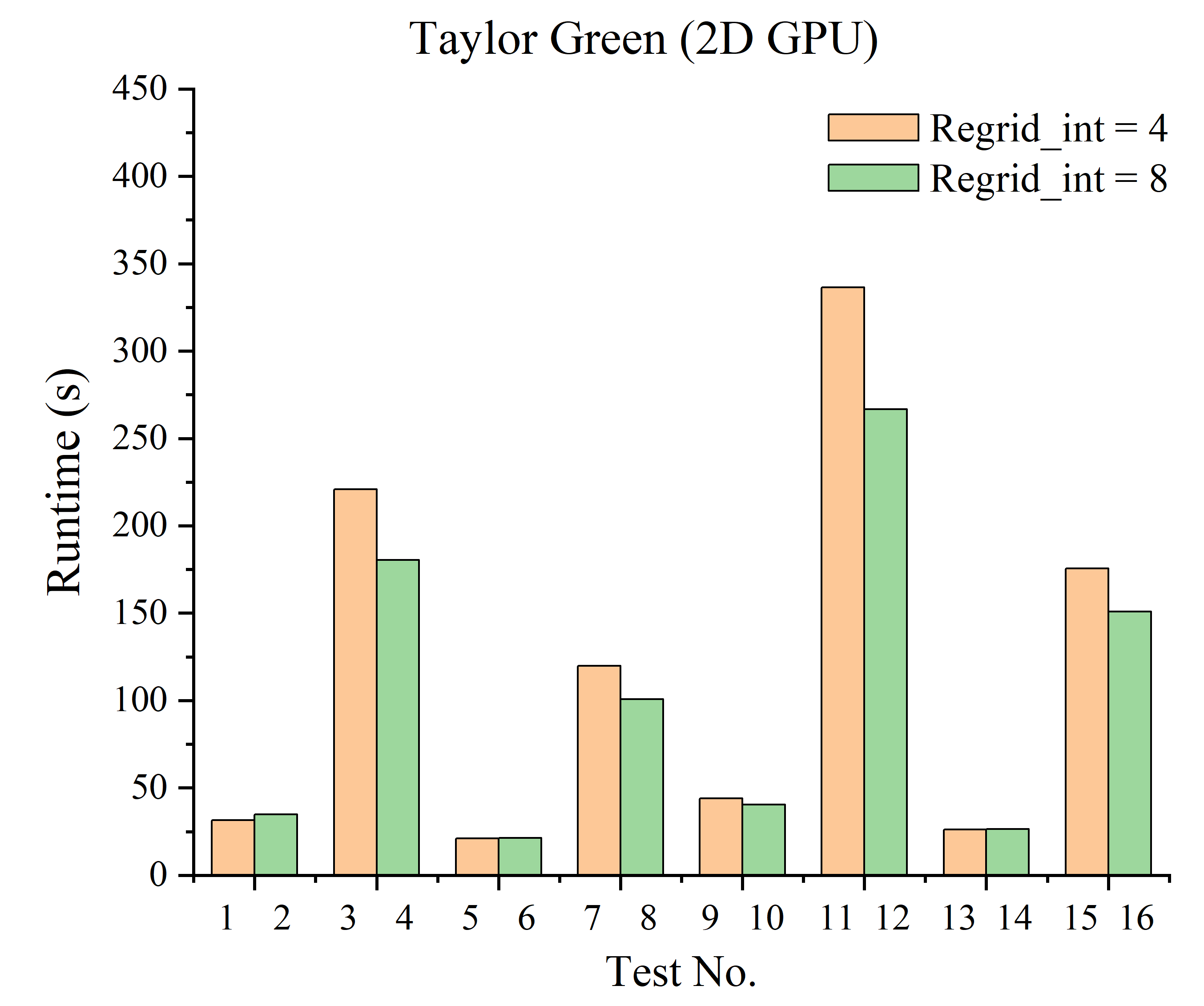}
  \caption{}
  \label{fig:regrid_int_d}
\end{subfigure}

\caption{Running time for various cases with different ${Regrid\_interval}$}
\label{fig:regrid_int}
\end{figure}

To investigate the influence of the parameter $Regrid\_interval$, Figure~\ref{fig:regrid_int} compares the running time of both Convected Vortex and Taylor Green Vortex cases with different values.~In various tests across different cases, the impact of $Regrid\_interval$ runtime varies, meaning that the runtime might increase or decrease with an increase in $Regrid\_interval$, necessitating a case-by-case analysis. An increase in $Regrid\_interval$ can lead to fewer instances of grid refinement and de-refinement, thereby reducing the additional overhead caused by grid interpolation and communication between different levels. On the flip side, a larger $Regrid\_interval$ may hinder the rapid capture of local feature changes in the flow field, potentially slowing down the convergence of the linear solvers. Additionally, larger refined patches might persist for a longer duration as the flow field progresses, potentially affecting the overall efficiency of the simulation.

\subsection{Skip\_level\_projection} \label{S:56}
\begin{figure}[H]
\centering
\begin{subfigure}{0.5\textwidth}
  \centering
  \includegraphics[width=\textwidth]{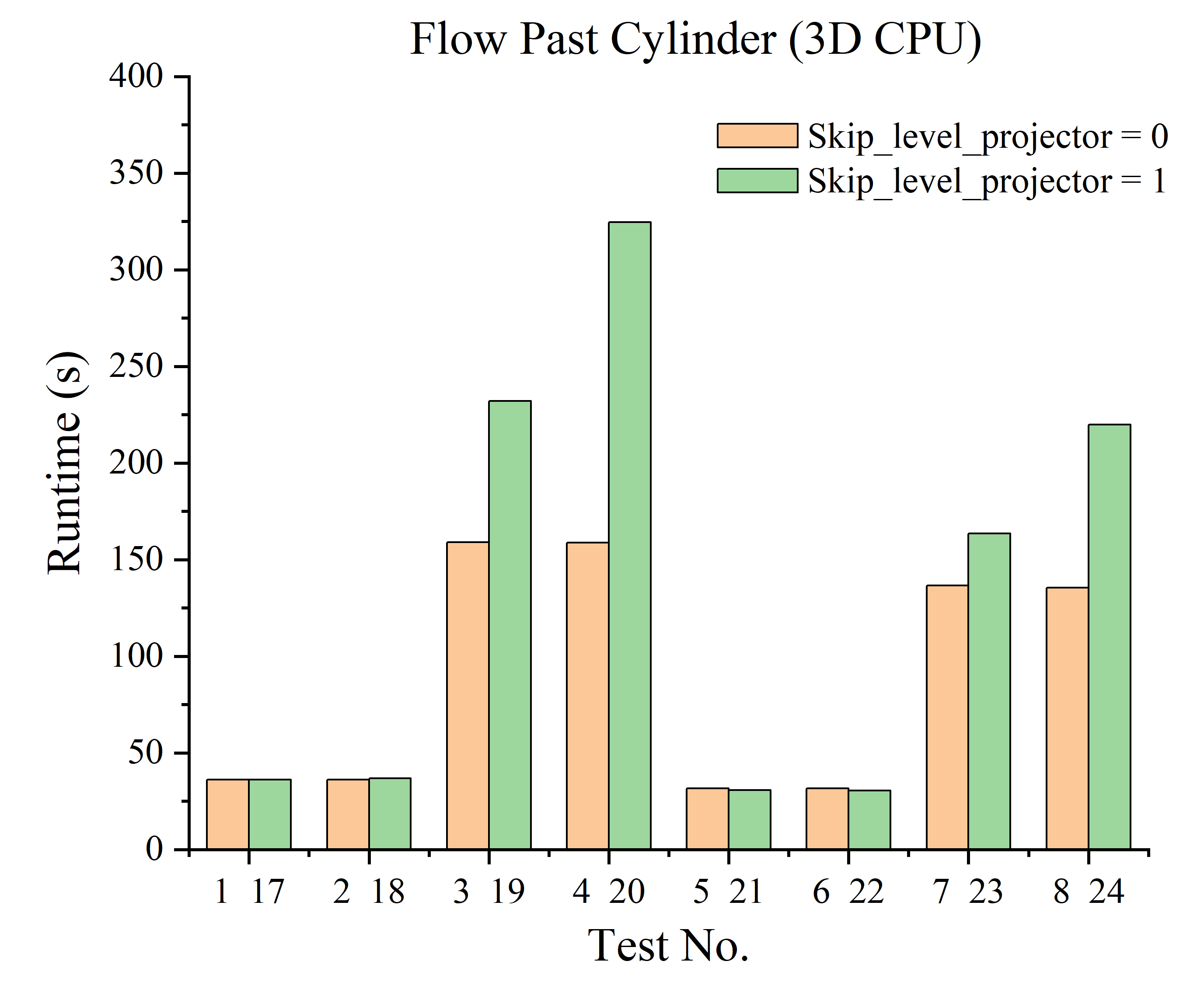}
  \caption{}
  \label{fig:skip_a}
\end{subfigure}\hfill
\begin{subfigure}{0.5\textwidth}
  \centering
  \includegraphics[width=\textwidth]{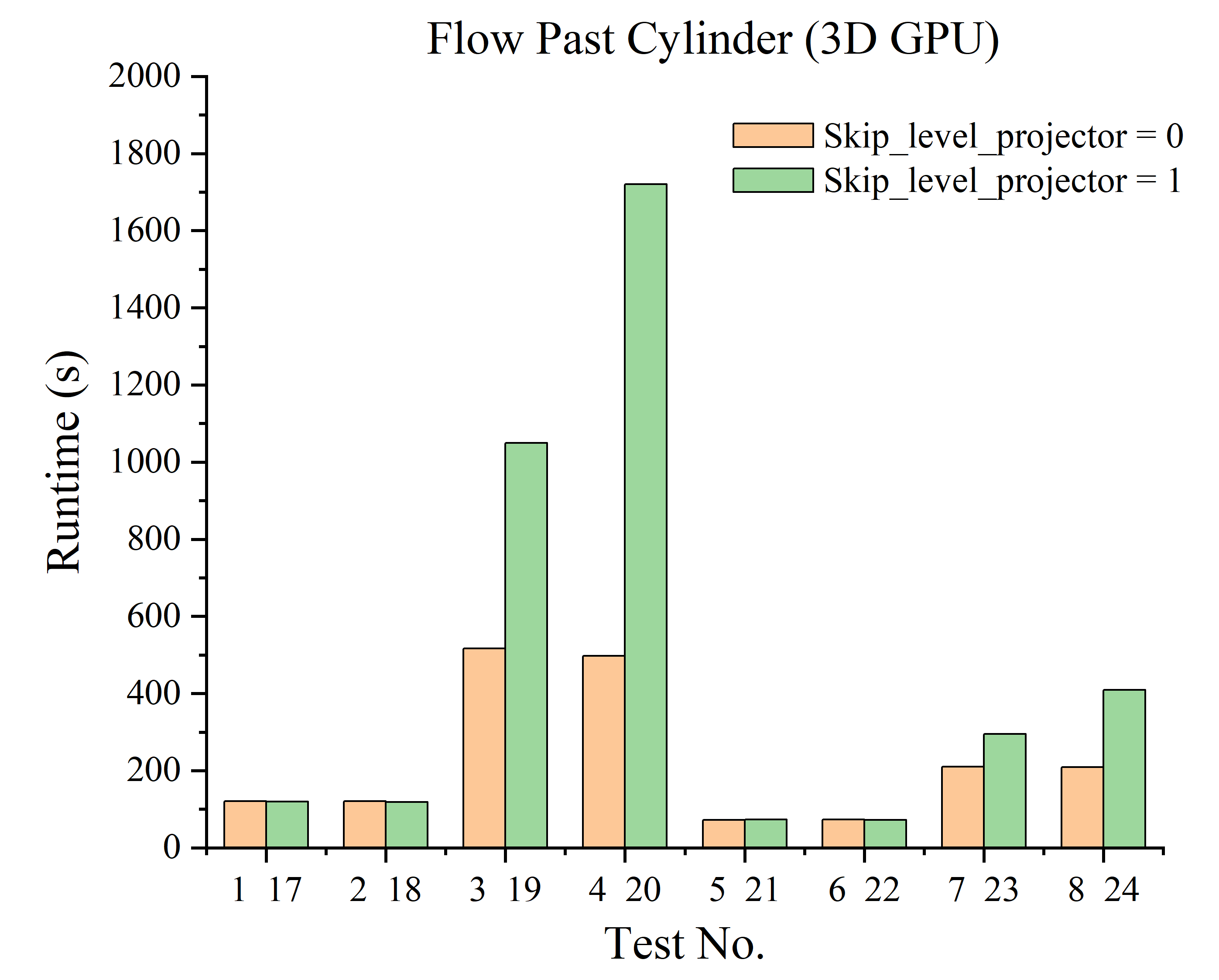}
  \caption{}
  \label{fig:skip_b}
\end{subfigure}

\caption{Running time for various cases with different ${Skip\_level\_projection}$}
\label{fig:skip_level_projection}
\end{figure}

The impact of $Skip\_level\_projection$ on runtime is generally minimal on all tests of a single case (Figure~\ref{fig:skip_level_projection}).~This may be attributed to the fact that the simulation time at coarser levels does not significantly contribute to the overall runtime. However, when $cycling =Auto$ and ${Max\_level=2}$, tests with $Skip\_level\_projection=1$ exhibit noticeably longer runtimes compared to those with $Skip\_level\_projection=0$. This could be due to the assumption behind skipping level projection—using the pressure values from the coarser grid at the previous timestep to approximate the current timestep’s pressure values—may not always be applicable in adaptive grid simulations where the grid changes dynamically.

\section{Conclusions and Future Directions} \label{S:6}

In this work, we explored the computational efficiency of incompressible flows using block-structured adaptive mesh refinement (BSAMR)~\citep{zhang2020amrex}. Our investigation sought to bridge this gap by conducting a comprehensive parametric study focusing on the implications of refining block size, refining frequency, maximum level of refinement, and the cycling method on computational time. Through extensive tests across CPUs and GPUs for diverse two-dimensional and three-dimensional cases, we have distilled several empirical insights. 

We found that a lower ${Max\_level}$ resulted in shorter runtime due to fewer grid numbers and a lower $Max\_grid\_size$ value led to longer runtime by increasing communication overhead between grid blocks. We then demonstrated that the choice between subcycling and non-subcycling methods depends on the simulation's goal. The subcycling method, which allows for different time steps on different levels, proved more efficient for achieving steady-state simulations when the time duration to reach a steady state is known. Conversely, for fixed-step simulations aimed at checking results, the non-subcycling method was recommended.

In addition, we analyzed the effects of $Regrid\_interva$l and $Skip\_level\_projection$ on runtime. It was found that the optimal setting for $Regrid\_interval$ depends on the specific case, as it can either increase or decrease runtime based on the balance between grid refinement overhead and the ability to capture flow field changes. Moreover, $Skip\_level\_projection$ had a negligible impact on runtime in most tests, except when it led to longer runtimes under certain conditions due to an assumption that might not hold in adaptive grid simulations.

For the CPU and GPU performance comparison, our study found that CPUs, with a specific configuration of processes and threads, generally outperformed GPUs except in cases with certain $Max\_grid\_size$ settings in three-dimensional simulations. This superior efficiency is attributed to GPUs' ability to better handle large $Max\_grid\_size$ values by reducing communication overhead and leveraging high-frequency CUDA cores for compute-intensive tasks.

In future research, we will further optimize and test the software parameters related to BSAMR for specific physical problems and automate this process to achieve maximum operational efficiency.

\section*{Acknowledgement}
The contributors of the IAMR open-source code are greatly acknowledged. 

\bibliographystyle{model1-num-names.bst}
\bibliography{refs}

\begin{thebibliography}{38}
\expandafter\ifx\csname natexlab\endcsname\relax\def\natexlab#1{#1}\fi
\providecommand{\url}[1]{\texttt{#1}}
\providecommand{\href}[2]{#2}
\providecommand{\path}[1]{#1}
\providecommand{\DOIprefix}{doi:}
\providecommand{\ArXivprefix}{arXiv:}
\providecommand{\URLprefix}{URL: }
\providecommand{\Pubmedprefix}{pmid:}
\providecommand{\doi}[1]{\href{http://dx.doi.org/#1}{\path{#1}}}
\providecommand{\Pubmed}[1]{\href{pmid:#1}{\path{#1}}}
\providecommand{\bibinfo}[2]{#2}
\ifx\xfnm\relax \def\xfnm[#1]{\unskip,\space#1}\fi
\bibitem[{Berger and Oliger(1984)}]{berger1984adaptive}
\bibinfo{author}{M.~J. Berger}, \bibinfo{author}{J.~Oliger},
\newblock \bibinfo{title}{{Adaptive mesh refinement for hyperbolic partial differential equations}},
\newblock \bibinfo{journal}{J. Comput. Phys.} \bibinfo{volume}{53} (\bibinfo{year}{1984}) \bibinfo{pages}{484--512}.
\bibitem[{Berger and Colella(1989)}]{berger1989local}
\bibinfo{author}{M.~J. Berger}, \bibinfo{author}{P.~Colella},
\newblock \bibinfo{title}{{Local adaptive mesh refinement for shock hydrodynamics}},
\newblock \bibinfo{journal}{J. Comput. Phys.} \bibinfo{volume}{82} (\bibinfo{year}{1989}) \bibinfo{pages}{64--84}.
\bibitem[{Griffith et~al.(2007)Griffith, Hornung, McQueen, and Peskin}]{griffith2007adaptive}
\bibinfo{author}{B.~E. Griffith}, \bibinfo{author}{R.~D. Hornung}, \bibinfo{author}{D.~M. McQueen}, \bibinfo{author}{C.~S. Peskin},
\newblock \bibinfo{title}{{An adaptive, formally second order accurate version of the immersed boundary method}},
\newblock \bibinfo{journal}{J. Comput. Phys.} \bibinfo{volume}{223} (\bibinfo{year}{2007}) \bibinfo{pages}{10--49}.
\bibitem[{Griffith(2012)}]{griffith2012immersed}
\bibinfo{author}{B.~E. Griffith},
\newblock \bibinfo{title}{{Immersed boundary model of aortic heart valve dynamics with physiological driving and loading conditions}},
\newblock \bibinfo{journal}{International journal for numerical methods in biomedical engineering} \bibinfo{volume}{28} (\bibinfo{year}{2012}) \bibinfo{pages}{317--345}.
\bibitem[{Bhalla et~al.(2013{\natexlab{a}})Bhalla, Bale, Griffith, and Patankar}]{bhalla2013unified}
\bibinfo{author}{A.~P.~S. Bhalla}, \bibinfo{author}{R.~Bale}, \bibinfo{author}{B.~E. Griffith}, \bibinfo{author}{N.~A. Patankar},
\newblock \bibinfo{title}{{A unified mathematical framework and an adaptive numerical method for fluid--structure interaction with rigid, deforming, and elastic bodies}},
\newblock \bibinfo{journal}{J. Comput. Phys.} \bibinfo{volume}{250} (\bibinfo{year}{2013}{\natexlab{a}}) \bibinfo{pages}{446--476}.
\bibitem[{Bhalla et~al.(2013{\natexlab{b}})Bhalla, Griffith, and Patankar}]{bhalla2013forced}
\bibinfo{author}{A.~P.~S. Bhalla}, \bibinfo{author}{B.~E. Griffith}, \bibinfo{author}{N.~A. Patankar},
\newblock \bibinfo{title}{A forced damped oscillation framework for undulatory swimming provides new insights into how propulsion arises in active and passive swimming},
\newblock \bibinfo{journal}{PLOS Comput. Biol.} \bibinfo{volume}{9} (\bibinfo{year}{2013}{\natexlab{b}}) \bibinfo{pages}{e1003097}.
\bibitem[{Zeng et~al.(2022)Zeng, Bhalla, and Shen}]{zeng2022subcycling}
\bibinfo{author}{Y.~Zeng}, \bibinfo{author}{A.~P.~S. Bhalla}, \bibinfo{author}{L.~Shen},
\newblock \bibinfo{title}{{A subcycling/non-subcycling time advancement scheme-based DLM immersed boundary method framework for solving single and multiphase fluid--structure interaction problems on dynamically adaptive grids}},
\newblock \bibinfo{journal}{Comput. Fluids}  (\bibinfo{year}{2022}) \bibinfo{pages}{105358}.
\bibitem[{Fujimoto and Machida(2006)}]{fujimoto2006electromagnetic}
\bibinfo{author}{K.~Fujimoto}, \bibinfo{author}{S.~Machida},
\newblock \bibinfo{title}{{Electromagnetic full particle code with adaptive mesh refinement technique: Application to the current sheet evolution}},
\newblock \bibinfo{journal}{Journal of Computational Physics} \bibinfo{volume}{214} (\bibinfo{year}{2006}) \bibinfo{pages}{550--566}.
\bibitem[{Balsara(2001)}]{balsara2001divergence}
\bibinfo{author}{D.~S. Balsara},
\newblock \bibinfo{title}{{Divergence-free adaptive mesh refinement for magnetohydrodynamics}},
\newblock \bibinfo{journal}{Journal of Computational Physics} \bibinfo{volume}{174} (\bibinfo{year}{2001}) \bibinfo{pages}{614--648}.
\bibitem[{Yao et~al.(2021)Yao, Jambunathan, Zeng, and Nonaka}]{yao2021massively}
\bibinfo{author}{Z.~Yao}, \bibinfo{author}{R.~Jambunathan}, \bibinfo{author}{Y.~Zeng}, \bibinfo{author}{A.~Nonaka},
\newblock \bibinfo{title}{{A massively parallel time-domain coupled electrodynamics--micromagnetics solver}},
\newblock \bibinfo{journal}{Int. J. High Perform. Comput. Appl.}  (\bibinfo{year}{2021}) \bibinfo{pages}{10943420211057906}.
\bibitem[{Santilli and Scotti(2015)}]{santilli2015stratified}
\bibinfo{author}{E.~Santilli}, \bibinfo{author}{A.~Scotti},
\newblock \bibinfo{title}{{The stratified ocean model with adaptive refinement (SOMAR)}},
\newblock \bibinfo{journal}{Journal of Computational Physics} \bibinfo{volume}{291} (\bibinfo{year}{2015}) \bibinfo{pages}{60--81}.
\bibitem[{van Hooft et~al.(2018)van Hooft, Popinet, van Heerwaarden, van~der Linden, de~Roode, and van~de Wiel}]{van2018towards}
\bibinfo{author}{J.~A. van Hooft}, \bibinfo{author}{S.~Popinet}, \bibinfo{author}{C.~C. van Heerwaarden}, \bibinfo{author}{S.~J. van~der Linden}, \bibinfo{author}{S.~R. de~Roode}, \bibinfo{author}{B.~J. van~de Wiel},
\newblock \bibinfo{title}{{Towards adaptive grids for atmospheric boundary-layer simulations}},
\newblock \bibinfo{journal}{Bound.-Layer Meteorol.} \bibinfo{volume}{167} (\bibinfo{year}{2018}) \bibinfo{pages}{421--443}.
\bibitem[{Sharma et~al.(2024)Sharma, Brazell, Vijayakumar, Ananthan, Cheung, deVelder, Henry~de Frahan, Matula, Mullowney, Rood et~al.}]{sharma2024exawind}
\bibinfo{author}{A.~Sharma}, \bibinfo{author}{M.~J. Brazell}, \bibinfo{author}{G.~Vijayakumar}, \bibinfo{author}{S.~Ananthan}, \bibinfo{author}{L.~Cheung}, \bibinfo{author}{N.~deVelder}, \bibinfo{author}{M.~T. Henry~de Frahan}, \bibinfo{author}{N.~Matula}, \bibinfo{author}{P.~Mullowney}, \bibinfo{author}{J.~Rood}, et~al.,
\newblock \bibinfo{title}{{ExaWind: Open-source CFD for hybrid-RANS/LES geometry-resolved wind turbine simulations in atmospheric flows}},
\newblock \bibinfo{journal}{Wind Energy}  (\bibinfo{year}{2024}).
\bibitem[{Mullowney et~al.(2021)Mullowney, Li, Thomas, Ananthan, Sharma, Rood, Williams, and Sprague}]{mullowney2021preparing}
\bibinfo{author}{P.~Mullowney}, \bibinfo{author}{R.~Li}, \bibinfo{author}{S.~Thomas}, \bibinfo{author}{S.~Ananthan}, \bibinfo{author}{A.~Sharma}, \bibinfo{author}{J.~S. Rood}, \bibinfo{author}{A.~B. Williams}, \bibinfo{author}{M.~A. Sprague},
\newblock \bibinfo{title}{Preparing an incompressible-flow fluid dynamics code for exascale-class wind energy simulations},
\newblock in: \bibinfo{booktitle}{Proceedings of the international conference for high performance computing, networking, storage and analysis}, \bibinfo{year}{2021}, pp. \bibinfo{pages}{1--16}.
\bibitem[{Khedkar et~al.(2021)Khedkar, Nangia, Thirumalaisamy, and Bhalla}]{khedkar2021inertial}
\bibinfo{author}{K.~Khedkar}, \bibinfo{author}{N.~Nangia}, \bibinfo{author}{R.~Thirumalaisamy}, \bibinfo{author}{A.~P.~S. Bhalla},
\newblock \bibinfo{title}{{The inertial sea wave energy converter (ISWEC) technology: device-physics, multiphase modeling and simulations}},
\newblock \bibinfo{journal}{Ocean Eng.} \bibinfo{volume}{229} (\bibinfo{year}{2021}) \bibinfo{pages}{108879}.
\bibitem[{Yu and Li(2013)}]{yu2013reynolds}
\bibinfo{author}{Y.-H. Yu}, \bibinfo{author}{Y.~Li},
\newblock \bibinfo{title}{{Reynolds-Averaged Navier--Stokes simulation of the heave performance of a two-body floating-point absorber wave energy system}},
\newblock \bibinfo{journal}{Comput. Fluids} \bibinfo{volume}{73} (\bibinfo{year}{2013}) \bibinfo{pages}{104--114}.
\bibitem[{Zeng and Shen(2020)}]{yadong2020osm}
\bibinfo{author}{Y.~Zeng}, \bibinfo{author}{L.~Shen},
\newblock \bibinfo{title}{{Modelling Wave Energy Converter (WEC) pointer absorbers using AMR techniques with both subcycling and non-subcycling}},
\newblock \bibinfo{journal}{Earth and Space Science Open Archive}  (\bibinfo{year}{2020}) \bibinfo{pages}{1}.
\bibitem[{Almgren et~al.(1998)Almgren, Bell, Colella, Howell, and Welcome}]{almgren1998conservative}
\bibinfo{author}{A.~S. Almgren}, \bibinfo{author}{J.~B. Bell}, \bibinfo{author}{P.~Colella}, \bibinfo{author}{L.~H. Howell}, \bibinfo{author}{M.~L. Welcome},
\newblock \bibinfo{title}{{A conservative adaptive projection method for the variable density incompressible Navier--Stokes equations}},
\newblock \bibinfo{journal}{J. Comput. Phys.} \bibinfo{volume}{142} (\bibinfo{year}{1998}) \bibinfo{pages}{1--46}.
\bibitem[{Popinet(2003)}]{popinet2003gerris}
\bibinfo{author}{S.~Popinet},
\newblock \bibinfo{title}{{Gerris: a tree-based adaptive solver for the incompressible Euler equations in complex geometries}},
\newblock \bibinfo{journal}{J. Comput. Phys.} \bibinfo{volume}{190} (\bibinfo{year}{2003}) \bibinfo{pages}{572--600}.
\bibitem[{Popinet(2009)}]{popinet2009accurate}
\bibinfo{author}{S.~Popinet},
\newblock \bibinfo{title}{{An accurate adaptive solver for surface-tension-driven interfacial flows}},
\newblock \bibinfo{journal}{J. Comput. Phys.} \bibinfo{volume}{228} (\bibinfo{year}{2009}) \bibinfo{pages}{5838--5866}.
\bibitem[{Williamschen and Groth(2013)}]{williamschen2013parallel}
\bibinfo{author}{M.~Williamschen}, \bibinfo{author}{C.~P. Groth},
\newblock \bibinfo{title}{{Parallel anisotropic block-based adaptive mesh refinement algorithm for three-dimensional flows}},
\newblock in: \bibinfo{booktitle}{21st AIAA Computational Fluid Dynamics Conference}, \bibinfo{year}{2013}, p. \bibinfo{pages}{2442}.
\bibitem[{Balaras and Vanella(2009)}]{balaras2009adaptive}
\bibinfo{author}{E.~Balaras}, \bibinfo{author}{M.~Vanella},
\newblock \bibinfo{title}{Adaptive mesh refinement strategies for immersed boundary methods},
\newblock in: \bibinfo{booktitle}{47th AIAA aerospace sciences meeting including the new horizons forum and aerospace exposition}, \bibinfo{year}{2009}, p. \bibinfo{pages}{162}.
\bibitem[{Chorin(1967)}]{chorin1967numerical}
\bibinfo{author}{A.~J. Chorin},
\newblock \bibinfo{title}{{The numerical solution of the Navier-Stokes equations for an incompressible fluid}},
\newblock \bibinfo{journal}{Bulletin of the American Mathematical Society} \bibinfo{volume}{73} (\bibinfo{year}{1967}) \bibinfo{pages}{928--931}.
\bibitem[{Sussman et~al.(1999)Sussman, Almgren, Bell, Colella, Howell, and Welcome}]{sussman1999adaptive}
\bibinfo{author}{M.~Sussman}, \bibinfo{author}{A.~S. Almgren}, \bibinfo{author}{J.~B. Bell}, \bibinfo{author}{P.~Colella}, \bibinfo{author}{L.~H. Howell}, \bibinfo{author}{M.~L. Welcome},
\newblock \bibinfo{title}{{An adaptive level set approach for incompressible two-phase flows}},
\newblock \bibinfo{journal}{J. Comput. Phys.} \bibinfo{volume}{148} (\bibinfo{year}{1999}) \bibinfo{pages}{81--124}.
\bibitem[{Sverdrup et~al.(2018)Sverdrup, Nikiforakis, and Almgren}]{sverdrup2018highly}
\bibinfo{author}{K.~Sverdrup}, \bibinfo{author}{N.~Nikiforakis}, \bibinfo{author}{A.~Almgren},
\newblock \bibinfo{title}{Highly parallelisable simulations of time-dependent viscoplastic fluid flow with structured adaptive mesh refinement},
\newblock \bibinfo{journal}{Phys. Fluids} \bibinfo{volume}{30} (\bibinfo{year}{2018}) \bibinfo{pages}{093102}.
\bibitem[{Zeng et~al.(2023)Zeng, Liu, Gao, Almgren, Bhalla, and Shen}]{zeng2023consistent}
\bibinfo{author}{Y.~Zeng}, \bibinfo{author}{H.~Liu}, \bibinfo{author}{Q.~Gao}, \bibinfo{author}{A.~Almgren}, \bibinfo{author}{A.~P.~S. Bhalla}, \bibinfo{author}{L.~Shen},
\newblock \bibinfo{title}{{A consistent adaptive level set framework for incompressible two-phase flows with high density ratios and high Reynolds numbers}},
\newblock \bibinfo{journal}{J. Comput. Phys.} \bibinfo{volume}{478} (\bibinfo{year}{2023}) \bibinfo{pages}{111971}.
\bibitem[{Zeng et~al.(2022)Zeng, Xuan, Blaschke, and Shen}]{zeng2022aparallel}
\bibinfo{author}{Y.~Zeng}, \bibinfo{author}{A.~Xuan}, \bibinfo{author}{J.~Blaschke}, \bibinfo{author}{L.~Shen},
\newblock \bibinfo{title}{A parallel cell-centered adaptive level set framework for efficient simulation of two-phase flows with subcycling and non-subcycling},
\newblock \bibinfo{journal}{J. Comput. Phys.} \bibinfo{volume}{448} (\bibinfo{year}{2022}) \bibinfo{pages}{110740}.
\bibitem[{Almgren et~al.(1996)Almgren, Bell, and Szymczak}]{almgren1996numerical}
\bibinfo{author}{A.~S. Almgren}, \bibinfo{author}{J.~B. Bell}, \bibinfo{author}{W.~G. Szymczak},
\newblock \bibinfo{title}{{A numerical method for the incompressible Navier--Stokes equations based on an approximate projection}},
\newblock \bibinfo{journal}{SIAM J. Sci. Comput.} \bibinfo{volume}{17} (\bibinfo{year}{1996}) \bibinfo{pages}{358--369}.
\bibitem[{Rider(1995)}]{rider1995approximate}
\bibinfo{author}{W.~J. Rider}, \bibinfo{title}{Approximate Projection Methods for Incompressible Flow: Implementation, Variants and Robustness}, \bibinfo{type}{LANL Unclassified Report} \bibinfo{number}{LA-UR-94-2000}, Los Alamos National Laboratory, \bibinfo{year}{1995}.
\bibitem[{Martin and Colella(2000)}]{martin2000cell}
\bibinfo{author}{D.~F. Martin}, \bibinfo{author}{P.~Colella},
\newblock \bibinfo{title}{{A cell-centered adaptive projection method for the incompressible Euler equations}},
\newblock \bibinfo{journal}{J. Comput. Phys.} \bibinfo{volume}{163} (\bibinfo{year}{2000}) \bibinfo{pages}{271--312}.
\bibitem[{Zeng et~al.(2022)Zeng, Xuan, Blaschke, and Shen}]{zeng2022parallel}
\bibinfo{author}{Y.~Zeng}, \bibinfo{author}{A.~Xuan}, \bibinfo{author}{J.~Blaschke}, \bibinfo{author}{L.~Shen},
\newblock \bibinfo{title}{{A parallel cell-centered adaptive level set framework for efficient simulation of two-phase flows with subcycling and non-subcycling}},
\newblock \bibinfo{journal}{J. Comput. Phys.} \bibinfo{volume}{448} (\bibinfo{year}{2022}) \bibinfo{pages}{110740}.
\bibitem[{Zeng and Shen(2019)}]{zeng2019unified}
\bibinfo{author}{Y.~Zeng}, \bibinfo{author}{L.~Shen},
\newblock \bibinfo{title}{{A unified AMR framework for multiphase flow and fluid-structure interaction problems with both non-subcycling and subcycling}},
\newblock in: \bibinfo{booktitle}{APS Division of Fluid Dynamics Meeting Abstracts}, \bibinfo{year}{2019}, pp. \bibinfo{pages}{S19--001}.
\bibitem[{Nonaka et~al.(2011)Nonaka, May, Almgren, and Bell}]{nonaka2011three}
\bibinfo{author}{A.~Nonaka}, \bibinfo{author}{S.~May}, \bibinfo{author}{A.~S. Almgren}, \bibinfo{author}{J.~B. Bell},
\newblock \bibinfo{title}{{A three-dimensional, unsplit godunov method for scalar conservation laws}},
\newblock \bibinfo{journal}{SIAM Journal on Scientific Computing} \bibinfo{volume}{33} (\bibinfo{year}{2011}) \bibinfo{pages}{2039--2062}.
\bibitem[{Bell et~al.(1991)Bell, Howell, and Colella}]{bell1991efficient}
\bibinfo{author}{J.~Bell}, \bibinfo{author}{L.~Howell}, \bibinfo{author}{P.~Colella},
\newblock \bibinfo{title}{{An efficient second-order projection method for viscous incompressible flow}},
\newblock in: \bibinfo{booktitle}{10th Computational Fluid Dynamics Conference}, \bibinfo{year}{1991}, p. \bibinfo{pages}{1560}.
\bibitem[{Zeng(2022)}]{zeng2022numerical}
\bibinfo{author}{Y.~Zeng}, \bibinfo{title}{Numerical Simulations of the Two-phase flow and Fluid-Structure Interaction Problems with Adaptive Mesh Refinement}, Ph.D. thesis, University of Minnesota, \bibinfo{year}{2022}.
\bibitem[{Zhang et~al.(2019)Zhang, Almgren, Beckner, Bell, Blaschke, Chan, Day, Friesen, Gott, Graves et~al.}]{zhang2019amrex}
\bibinfo{author}{W.~Zhang}, \bibinfo{author}{A.~Almgren}, \bibinfo{author}{V.~Beckner}, \bibinfo{author}{J.~Bell}, \bibinfo{author}{J.~Blaschke}, \bibinfo{author}{C.~Chan}, \bibinfo{author}{M.~Day}, \bibinfo{author}{B.~Friesen}, \bibinfo{author}{K.~Gott}, \bibinfo{author}{D.~Graves}, et~al.,
\newblock \bibinfo{title}{{AMReX: a framework for block-structured adaptive mesh refinement}},
\newblock \bibinfo{journal}{J. Open Source Softw.} \bibinfo{volume}{4} (\bibinfo{year}{2019}).
\bibitem[{Zhang et~al.(2020)Zhang, Myers, Gott, Almgren, and Bell}]{zhang2020amrex}
\bibinfo{author}{W.~Zhang}, \bibinfo{author}{A.~Myers}, \bibinfo{author}{K.~Gott}, \bibinfo{author}{A.~Almgren}, \bibinfo{author}{J.~Bell},
\newblock \bibinfo{title}{{AMReX: Block-Structured Adaptive Mesh Refinement for Multiphysics Applications}},
\newblock \bibinfo{journal}{arXiv preprint arXiv:2009.12009}  (\bibinfo{year}{2020}).
\bibitem[{Zeng et~al.(2021)Zeng, Bhalla, He, and Shen}]{zeng2021subcycling}
\bibinfo{author}{Y.~Zeng}, \bibinfo{author}{A.~P. Bhalla}, \bibinfo{author}{S.~He}, \bibinfo{author}{L.~Shen},
\newblock \bibinfo{title}{A subcycling/non-subcycling time advancement scheme-based sharp-interface immersed boundary method framework for solving fluid-structure interaction problems on dynamically adaptive grids},
\newblock in: \bibinfo{booktitle}{APS Division of Fluid Dynamics Meeting Abstracts}, \bibinfo{year}{2021}, pp. \bibinfo{pages}{F26--004}.

\end{thebibliography}

\end{document}